\documentclass{aa}
	
\usepackage[varg]{txfonts}
\usepackage{natbib,twoopt}
\usepackage[breaklinks=true]{hyperref} %% to avoid \citeads line fills
\bibpunct{(}{)}{;}{a}{}{,} %% natbib format for A&A and ApJ
\makeatletter
\newcommandtwoopt{\citeads}[3][][]{\href{http://adsabs.harvard.edu/abs/#3}%
{\def\hyper@linkstart##1##2{}%
\let\hyper@linkend\@empty\citealp[#1][#2]{#3}}}
\newcommandtwoopt{\citepads}[3][][]{\href{http://adsabs.harvard.edu/abs/#3}%
{\def\hyper@linkstart##1##2{}%
\let\hyper@linkend\@empty\citep[#1][#2]{#3}}}
\newcommandtwoopt{\citetads}[3][][]{\href{http://adsabs.harvard.edu/abs/#3}%
{\def\hyper@linkstart##1##2{}%
\let\hyper@linkend\@empty\citet[#1][#2]{#3}}}
\newcommandtwoopt{\citeyearads}[3][][]%
{\href{http://adsabs.harvard.edu/abs/#3}
{\def\hyper@linkstart##1##2{}%
\let\hyper@linkend\@empty\citeyear[#1][#2]{#3}}}
\makeatother
\usepackage{graphicx} % figuras
\usepackage{multicol}
\usepackage{subfig}
\usepackage[export]{adjustbox}
\usepackage{calligra}
\DeclareMathAlphabet{\mathcalligra}{T1}{calligra}{m}{n}
\DeclareFontShape{T1}{calligra}{m}{n}{<->s*[2.2]callig15}{}

\begin{document}
\title{Revisiting neutron starquakes caused by spin-down}
\author{\textbf{Javier A. Rencoret}\inst{1}, \textbf{Claudia Aguilera-G\'omez}\inst{2}
\and \textbf{Andreas Reisenegger}$^{3}$ 
\offprints{J.A.Rencoret, \email{jsrencoret@gmail.com}}}

\institute{Instituto de F\'isica, Pontificia Universidad Cat\'olica de Valpara\'iso, Av. Universidad 330, Curauma, Valpara\'iso, Chile \ 
\and N\'ucleo de Astronom\'ia, Universidad Diego Portales, Ej\'ercito 441, Santiago, Chile
\and Departamento de F\'{\i}sica, Facultad de Ciencias B\'asicas, Universidad Metropolitana de Ciencias de la Educaci\'on, Av. Jos\'e Pedro Alessandri 774, \~Nu\~noa, Santiago, Chile}
\titlerunning{Neutron starquakes}
\authorrunning{Rencoret et al.}

%\date{Received  / Accepted }

\abstract{Pulsars show a steady decrease in their rotational frequency, occasionally interrupted by sudden spin-ups called glitches, whose physical origin is still a mystery. One suggested explanation for at least the small glitches are starquakes, that is, failures of the solid neutron star crust, in which the progressive reduction in the centrifugal force deforms the star, stressing the solid until it breaks. This produces a spin-up, dissipating energy inside the star.}
{We explore this suggestion by analyzing a mostly analytical model in order to understand the possible consequences of starquakes, particularly whether they can explain at least the small glitches.}
{We analyze the deformations and strains produced by the decreasing centrifugal force, modeling the neutron star with a fluid core and a solid crust, each with uniform density and with the core possibly denser than the crust, as a simple approximation to the strong density gradient present in real neutron stars.} 
{The deformation of a star with very different densities in the core and crust is qualitatively different from the previously studied case of equal densities. The former more closely resembles the behavior of a fluid star, in which the core-crust interface is a surface of constant gravitational plus centrifugal potential.} 
{Regardless of the uncertain breaking strain, the glitch activity in this model is several orders of magnitude smaller than observed, even if only small glitches are considered. For a large breaking strain, suggested by simulations, glitches due to starquakes could be roughly of the correct size but much less frequent than observed glitches. The energy released in each such glitch is much larger than in the standard model of angular momentum transfer from a faster rotating superfluid in the inner crust.  On the other hand, we cannot rule out that the heating produced by small starquakes could trigger glitches by allowing neutron superfluid vortices to move. We also confirm that stresses in the neutron star crust can in principle support an ellipticity much larger than some observational upper limits from pulsar timing and continuous gravitational wave searches.}


\maketitle

\section{Introduction}
Pulsars are among the best clocks known in the universe, with very precisely periodic pulses due to the rotation of a neutron star with a strong, inclined dipolar magnetic field. These pulses slow down progressively as the star loses rotational energy. In addition, they exhibit timing irregularities, the most prominent of which are "glitches", sudden spin-up events, in which about $1\,\%$ of the secular spin-down is recovered \citep{2000MNRAS.315..534L,2011MNRAS.414.1679E,2017A&A...608A.131F}.

Since the detection of the first glitches \citep{1968Natur.217..709H,1969IAUC.2179....1B}, the scientific community has tried to understand their physical mechanism. Initially, \citet{1969Natur.223..597R} and \citet{1969Natur.224..872B} explained them as "starquakes", sudden fractures of the elastic neutron star crust.
These authors argued that because of its fast rotation, the pulsar has an ellipticity $\epsilon_{0}$, which, as the star spins down, would change toward a more spherical shape by $\Delta\epsilon_0$ if the star were entirely fluid. In a neutron star, only the core is fluid, while the crust is an elastic solid; therefore, the crust resists the deformation and accumulates stress, reducing the change in ellipticity by $\Delta\epsilon=b\Delta\epsilon_{0}$, where $b$ is the (very small) "rigidity parameter" of the neutron star (e.g., \citealt{2003ApJ...588..975C}). When a critical strain is reached, the crust yields and relaxes, so the star becomes more spherical by an amount $\Delta\epsilon$, decreasing its moment of inertia and thus causing a sudden increase in the rotation rate. 

In this scenario, the time between starquakes is proportional to the value of $\Delta\epsilon_0=\Delta\epsilon_0^{max}$ at which the crust breaks (set by its critical strain angle), and the size of the associated glitches is proportional to the corresponding value of $\Delta\epsilon^{max}=b\Delta\epsilon_0^{max}$ \citep{1971AnPhy..66..816B}. Thus, %we note that 
the ``glitch activity,'' $\dot\nu_g$, the long-term average spin-up rate due to glitches  \citep{2000MNRAS.315..534L}, is independent of the rather uncertain critical strain of the neutron star crust but proportional to the rigidity parameter $b$. 

\citet{1971AnPhy..66..816B} formulated the starquake problem by writing the total energy of the rotating, stressed star as
\begin{equation}
\label{energy B&P}
    E=E_0+\frac{L^2}{2I}+A\epsilon^2+B(\epsilon-\epsilon_0)^2,
\end{equation}
where $E_0$ is the energy of the relaxed, nonrotating star, $L$ and $I$ are the angular momentum and moment of inertia, the third term is the correction to the gravitational potential energy due to the rotation-induced ellipticity $\epsilon$, and the last term accounts for the elastic stress in the crust, where $\epsilon_0$ is an ellipticity at which the crust is relaxed. They evaluated the coefficients $A$ and $B$ as integrals over the displacement field for a completely solid, incompressible star with uniform density, uniform shear modulus, and Newtonian gravity (later reformulated by \citealt{1975AnPhy..95...74C} in general relativity). The presence of the fluid core was taken into account in a nonrigorous, heuristic fashion by restricting the integral for $B$ to the volume of the crust, obtaining the rigidity parameter $b=B/(A+B)\approx 10^{-5}$ for a star of crustal shear modulus $\mu=10^{30}\,\mathrm{erg\,cm^{-3}}$ and mass $M=1.36\,M_\odot$ (and increasing toward smaller masses, for which the gravity is weaker and the crust is thicker). 

At that point, with roughly one year of timing observations, only two glitches had been detected, namely a very small one in the Crab and a substantially larger one in the Vela pulsar, %whereas 
the latter of which spins down more slowly than the former. Assuming that these glitches are ``typical'' for each of these pulsars and occur roughly once per year, \citet{1971AnPhy..66..816B} concluded that starquakes could at best account for the small Crab glitch but by no means for the much larger glitch of Vela, unless it was an extremely unusual event. As a more viable alternative, \citet{1975Natur.256...25A} explained glitches as sudden transfers of angular momentum from a weakly coupled neutron superfluid in the inner crust to the rest of the star. Later, the detection of many more glitches in these and many other pulsars allowed us to infer that their size distribution is (at least) bimodal \citep{2000MNRAS.315..534L,2011MNRAS.414.1679E,2017A&A...608A.131F}, suggesting that different mechanisms might be responsible for small and large glitches. For instance, angular momentum transfer from a superfluid component could be invoked for the large glitches \citep{2011ApJ...743L..20P,2015MNRAS.449.3559H,2016MNRAS.460.1201H}, perhaps still leaving room for starquakes to potentially explain the small ones (as mentioned, e.g., by \citealt{2017PhRvL.118z1101J}).

On the other hand, \citet{1998ApJ...508..838L} and \citet{2000ApJ...543..987F} assumed that glitches are caused by starquakes that fracture the crust asymmetrically (biased by the structure of the stellar magnetic field), causing a damped precessional motion of the star, which would result in an increase in the angle between the rotational and magnetic axes and thus explain the observation that the spin-down rate is increased by each glitch. The first of these papers used the model of \citet{1971AnPhy..66..816B}, whereas the second improved on it, calculating the displacement field for a star with a fluid core and an elastic crust (still incompressible, with uniform density everywhere in the star, uniform shear modulus in the crust, and Newtonian gravity). From this model, one obtains a five times smaller rigidity parameter $b$ (for the same assumed shear modulus). 

The discovery of correlated, quasi-periodic variations in the pulse shape and spin-down rate of PSR B1828$-$11 and its interpretation as free precession by \citet{2000Natur.406..484S} motivated \citet{2003ApJ...588..975C} to consider a more realistic (though still Newtonian) model (following \citealt{2000MNRAS.319..902U}) with compressible matter whose density and shear-modulus profiles are obtained from equations of state based on nuclear physics calculations, yielding an even smaller value, $b\approx 2\times 10^{-7}$. The dependence on the assumed neutron star mass and equation of state was explored by \citet{2008A&A...491..489Z} (using an approximate, heuristic formula for $B$ based on the results of \citealt{2003ApJ...588..975C}), generally finding similar values (with a spread of a factor of a few). 

The possibility of detecting continuous gravitational waves from neutron stars with an elastically deformed crust motivated further studies, such as that of \citet{2013PhRvD..88d4004J}, who calculated the maximum quadrupole moment sustainable by a stressed crust (as well as other, more exotic alternatives) in general relativity, finding that it is $\sim 6$ times smaller than the Newtonian result, a correction that should also apply to the closely related rigidity parameter $b$.

Recently, \citet{2020MNRAS.491.1064G} numerically calculated the centrifugal deformation of a (Newtonian) polytropic star with different adiabatic indices and confirmed that although the value of the adiabatic index has some effect on the deformations, the strain accumulated between successive glitches of the Vela pulsar is in all cases much too small to break the crust; therefore, if starquakes are invoked to trigger glitches, the crust must always remain close to the breaking strain. The same authors \citep{2019PASA...36...36G} analytically calculated the centrifugal deformation for a star with uniform but possibly different densities in the fluid core and the elastic crust. Among other things, they analyzed the effect of the Cowling approximation (neglecting perturbations of the gravitational potential; \citealt{1941MNRAS.101..367C}) for the case of equal densities in the crust and the core, finding that the displacements with this approximation are $\approx 2/5$ of those for the exact solution. They do not evaluate the rigidity parameter $b$ or the angular velocity changes resulting from eventual starquakes, which we do in the present paper.

Despite these theoretical advances and the much increased sample of glitches (now numbering $\sim 400$; e.g. \citealt{2017A&A...608A.131F}), we are not aware of any papers reanalyzing the question first raised (and to some extent answered) by \citet{1971AnPhy..66..816B}, namely whether starquakes are large and frequent enough to explain all the observed glitches or even a subset of these, such as the class of small glitches. 

In the present paper, we consider the same model as \citet{2019PASA...36...36G}, namely a Newtonian star with uniform but possibly different densities in the fluid core and the solid crust,\footnote{Like these and, to our knowledge, all previous authors, we ignore that as the pressure at the core-crust interface is perturbed, some matter elements might melt or solidify, so the displacement of the interface is somewhat different from that of the matter elements adjacent to it.}  as a simple approximation to the strong (but continuous) density stratification expected to be present in real neutron stars. In order not to overcomplicate this already quite difficult problem, we apply the Cowling approximation, neglecting the gravitational potential perturbations in our calculations but making an approximate correction for it at the end. This approach allows us to analyze the displacements caused by the centrifugal force in the fluid core and the elastic crust of the star, and to show that there is a competition between the buoyancy force associated with the core-crust density difference (which tends to make the crust-core interface an equipotential surface for the combined gravitational plus centrifugal potential) and the shear stress (which tends to minimize its deformation from the original, unstressed shape).

We go on to show that although starquakes can qualitatively resemble glitches, the glitch activity associated with them is far too small to explain even the subclass of small glitches. The predicted glitch activity, and thus this conclusion, are independent of the assumed critical breaking strain $\sigma_c$, since the size of the glitches is proportional to $\sigma_c$, whereas their frequency is $\propto 1/\sigma_c$. 

On the other hand, for the large yield strains found in molecular dynamics simulations of neutron star crust matter \citep{2009PhRvL.102s1102H}, we find that spin-down can produce starquakes at most a few times in the life of a neutron star. We also analyze the temperature increase that could be caused by starquakes, which might allow neutron star vortices to move, thus triggering glitches. Finally, we use the previous results to infer the possible ellipticity that might be sustained by crustal stresses in our model, confirming that it is much larger than observational bounds established for millisecond pulsars.

This article is organized as follows. Section \ref{sec:deformations} describes the displacement field caused in the star by a changing centrifugal force, we discuss the differences between fluid and elastic matter and evaluate the case of a fluid core and an elastic crust as a function of their density difference and the shear modulus of the crust. Section \ref{sec:glitch} shows that the predicted glitch activity due to starquakes is much lower than that observed even for small glitches, and we use estimates for the breaking strain to predict the likely size and frequency of glitches caused by starquakes. Other potentially observable effects of starquakes are discussed in Sect. \ref{sec:energy released} . Section \ref{sec:ellipticity} evaluates the ellipticity that can be sustained by crustal stresses in our model and compare it to observational constraints based on gravitational-wave emission from millisecond pulsars. Section \ref{sec:conclusions} contains the conclusions of this work.

\section{Rotational deformations of a neutron star}
\label{sec:deformations}

\begin{figure}
\centering
\resizebox{\hsize}{!}{\includegraphics{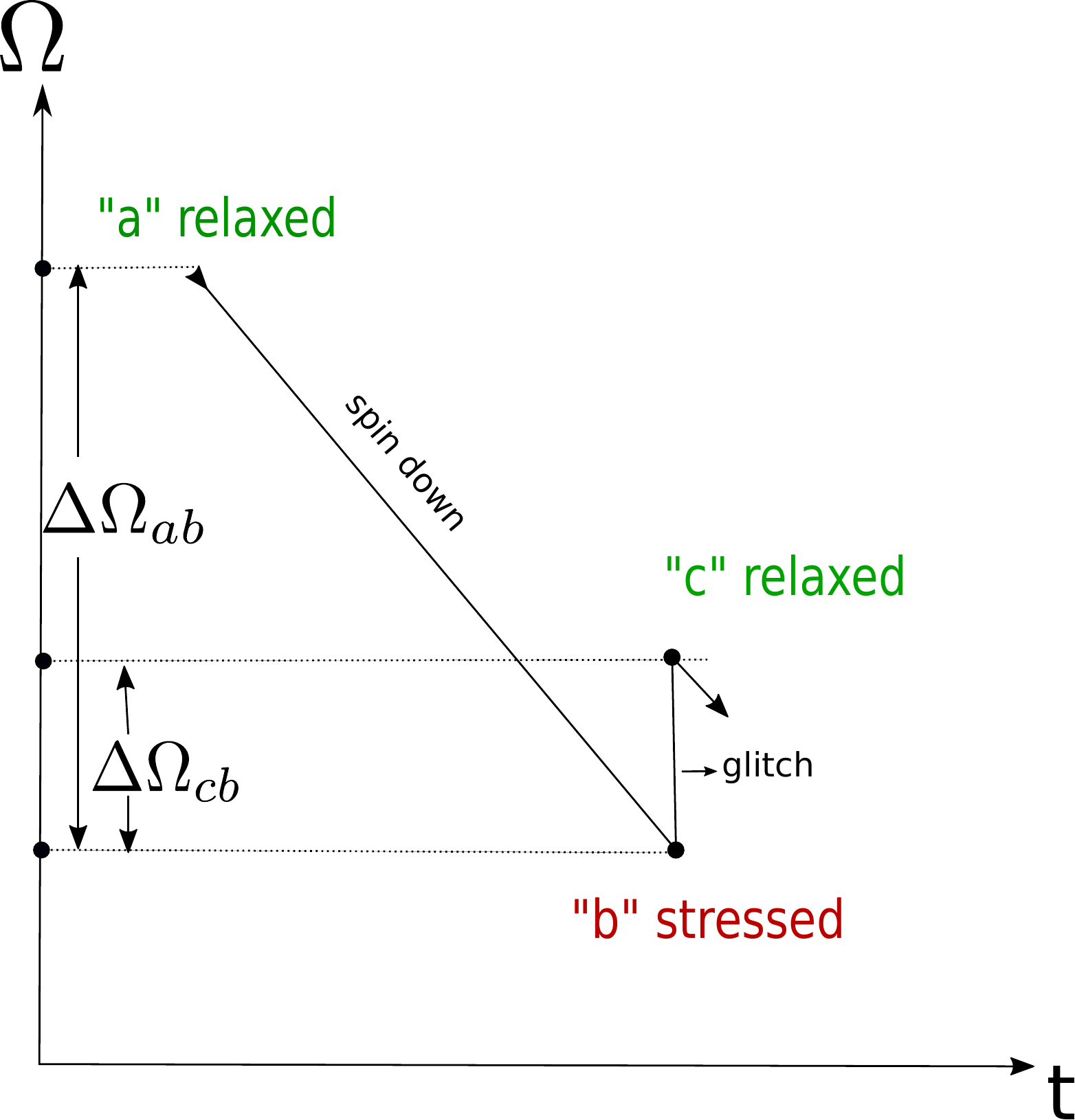}}\\
\caption{Schematic diagram of a starquake driven by spin-down, with time $t$ on the horizontal axis and rotation rate $\Omega$ on the vertical axis. In state "a", the star is relaxed. From "a" to "b" it spins down, with its crust deforming and building up stress. At "b", the crust reaches a critical stress and fractures, the moment of inertia decreases, and the star suddenly spins up to a new relaxed state "c".} \label{fig:esquemastarquake.pdf}
\end{figure}

As explained above, in the starquake scenario the decreasing centrifugal force deforms the neutron star toward a more spherical shape, stressing the elastic crust, which eventually yields, allowing an abrupt, additional deformation that reduces the moment of inertia and thus spins up the star. In an extreme quake (which we consider for simplicity), all elastic stresses are released in the quake. Thus, if, as illustrated in Fig.~\ref{fig:esquemastarquake.pdf}, the star starts from a relaxed state ``a'' (at birth or after a previous, large quake), we can regard the state ``b'' just before and the state ``c'' just after the quake as perturbations of ``a'' under the same applied force (the change in the centrifugal force), in which the process a $\to$ b is subject to the elastic stresses in the crust but a $\to$ c is not, that is, in the latter process the star behaves as if completely fluid. Therefore, we need to analyze the displacement fields characterizing the deformation of both an elastic solid and a fluid under a changing force field. For conceptual and notational clarity, we start by introducing the basic concepts and equations of elastostatics, followed by the calculation of the displacement fields for the elastic and the fluid matter.

\subsection{Equilibrium elastic deformation under an arbitrary force}
\label{sec:formalism}

The elastostatic equilibrium condition %for a neutron star 
in Newtonian gravity is
\begin{equation}
\label{elastoestatic equilibrium}
-\vec{\nabla}\cdot \tens{T}+\rho \vec{g}+\vec{f}_c=0
\end{equation}
(e.g., \citealt{1999PhRvL..83.3362L, 2003ApJ...588..975C, thorne2017modern}), where $\tens{T}$ is the stress tensor\footnote{For the stress tensor, we follow the sign convention of \citet{thorne2017modern}, which is opposite to that of \citet{2000ApJ...543..987F} and \citet{2003ApJ...588..975C}.}, $\rho$ is the mass density, $\vec{g}=-\nabla\phi_g$ is the acceleration of gravity ($\phi_g$ is the Newtonian gravitational potential), and $\vec{f}_c$ is some other force density field acting on the stellar matter (which we later take to be the centrifugal force due to the stellar rotation but which could, in principle, represent other forces, such as a Lorentz force of the stellar magnetic field). The stress tensor can be decomposed as 
\begin{equation}
\label{stress tensor}
\tens{T}=p\tens{g}+\tens{S},
\end{equation}
where $p$ is the pressure, $\tens{g}$ is the metric tensor of Euclidean, three-dimensional space (not to be confused with the gravitational acceleration $\vec{g}$ defined above), and the symmetric, trace-free tensor $\tens{S}$ represents the shear stress, which vanishes in a fluid and in a relaxed solid. 
With this decomposition, Eq.~(\ref{elastoestatic equilibrium}) can be rewritten as
\begin{equation}
\label{elastoestatic equilibrium 2}
-\vec{\nabla}p+\vec{f}_{sh}-\rho\vec{\nabla}\phi_g+\vec{f}_c=0,
\end{equation}
where 
\begin{equation}
\label{shear force}
    \vec{f}_{sh}=-\vec{\nabla}\cdot\tens{S}
\end{equation} 
is the elastic force density produced by the shear stress. 

If we consider a stressed state "b" (with $\vec{f}_{sh}\neq 0$) close to an unstressed state "a" (with $\vec{f}_{sh}=0$), the small Eulerian (local) changes of the variables (denoted by $\delta[.]\equiv[.]_b-[.]_a$) are related by 
\begin{equation}
\label{elastoestatic change 2}
-\vec{\nabla}\delta p +\vec{f}_{sh}-\delta\rho\vec{\nabla}\phi_g -\rho\vec{\nabla}\delta\phi_g +\delta\vec{f}_c=0.
\end{equation}
We also need to consider Lagrangian (material) perturbations, which give the change in a physical quantity in a given matter element as it is displaced, e.g., $\Delta T^{ij}=T^{ij}_{b}(\vec{x}+\vec{\xi})-T^{ij}_{a}(\vec{x})$, where $\vec{\xi}$ represents the displacement vector field from the position $\vec{x}$ of a matter element in state "a" to its position $\vec{x}+\vec{\xi}$ in state "b".
In the linear limit, Eulerian and Lagrangian perturbations of any fluid variable are related by
\begin{equation}
\Delta(.)=\delta(.)+\vec\xi\cdot\nabla(.).
\end{equation}

In the elastic limit, the Lagrangian stress perturbation is proportional to the strain tensor (generalized Hooke's law)\footnote{Throughout this paper, we express tensor components in a coordinate basis \citep{Schutz09} (in the following sections specifically spherical coordinates), distinguishing between ``covariant'' and ``contravariant'' components (with indices as subscripts or superscripts), and interpreting $\nabla_k$ as a covariant derivative with respect to the coordinate $x^k$. We also use the ``Einstein convention'', implicitly assuming a sum over indices that appear as subscripts and superscripts in the same expression, here $k$ and $l$.},
\begin{equation}
\label{Hooke law}
\Delta T^{ij}=-Y^{ijkl}\sigma_{kl},
\end{equation} 
where the rank-four tensor $Y^{ijkl}$ is known as the Young tensor (a property of the material) and 
\begin{equation}
\sigma_{kl}=\frac{1}{2}\left(\nabla_{k}\xi_{l}+\nabla_{l}\xi_{k}\right)
\end{equation}
is the strain tensor, which can be decomposed into its irreducible tensorial parts as
\begin{equation}
\label{strain tensor}
\tens{\sigma}=\frac{1}{3}\Theta\tens{g}+\tens{\Sigma}+\tens{R},
\end{equation}
where the trace
\begin{equation}
\label{expansion}
\Theta=\vec{\nabla}\cdot\vec{\xi}
\end{equation}
represents the expansion (or compression) of the matter element, the symmetric, traceless part
\begin{equation}
\label{shear strain}
\Sigma^{ij}=\frac{1}{2}(\nabla^{i}\xi^{j}+\nabla^{j}\xi^{i})-\frac{1}{3}\Theta g^{ij}
\end{equation}
represents the shear deformations, and the antisymmetric part \begin{equation}
\label{rotation}
R^{ij}=\frac{1}{2}(\nabla^{i}\xi^{j}-\nabla^{j}\xi^{i})
\end{equation}
represents rotations (see \citealt{thorne2017modern} for a pedagogical discussion).

In an isotropic solid, Hooke's law (Eq.~(\ref{Hooke law})) can be written separately for the two pieces of the stress tensor (Eq.~(\ref{stress tensor})) as 
\begin{equation}
\label{Hooke pressure}
\Delta p=-K\Theta
\end{equation}
and 
\begin{equation}
\label{Hooke shear}
    \tens{S}=-2\mu\tens{\Sigma},
\end{equation} 

where $K$ is the bulk modulus, which quantifies the resistance of the matter to expansion or compression, and $\mu$ is the shear modulus, which accounts for the resistance of the solid to shear deformations (rotations do not produce stress; \citealt{thorne2017modern}). Thus, the shear force in Eqs. (\ref{elastoestatic equilibrium 2}) and (\ref{elastoestatic change 2}) can be written as 
\begin{equation}
    \vec{f}_{sh}=2\vec{\nabla}\cdot(\mu\tens{\Sigma}),
\end{equation}
with $\tens\Sigma$ given by Eq. (\ref{shear strain}).

In order to obtain the displacement field $\vec\xi(\vec{x})$ caused by the change $\delta\vec{f}_c(\vec{x})$ in the applied force field, we must solve Eq. (\ref{elastoestatic change 2}) with the appropriate boundary conditions, to be discussed in Sect.~\ref{sec:boundary}.

\subsection{Elastic deformations in a rotating star}
\label{sec:elastic}

For mathematical simplicity and to allow physical insight, we consider a simple model in which the neutron star matter is incompressible ($K\to\infty$), with $\Theta=0=\Delta\rho$ (but $\Delta p=-K\Theta\neq 0$). We also assume uniform density in each domain (core and crust), so the Eulerian density perturbation $\delta\rho$ vanishes as well (except at the boundaries). We also consider a uniform shear modulus $\mu$ throughout the crust (and of course $\mu=0$ in the fluid core). Up to this point, the model is identical to that of \citet{2000ApJ...543..987F}. However, as recently done by \citet{2019PASA...36...36G}, we allow for different densities in the two domains, $\rho_s$ in the elastic, solid crust, and $\rho_f=\rho_s(1+\Delta\tilde\rho)$ in the fluid core, which it makes an important difference.

When the star spins down, it is deformed (toward a progressively more spherical shape) by the change in the centrifugal force density, $\delta\vec{f}_c=-\rho\vec{\nabla}\delta\phi_c$. 
Since we take $\rho$ to be uniform and $\delta\rho=0$, Eq. (\ref{elastoestatic change 2}) implies that the shear force can be derived from a potential,
\begin{equation}
\label{elastoestatic change 3}
\vec{f}_{sh}=\vec{\nabla}\left(\delta p+\rho\delta\phi_T\right)
\equiv-\vec{\nabla}\Psi_{sh},
\end{equation}
where we introduced a total (effective) potential, $\phi_T\equiv\phi_g+\phi_c$. Thus,
\begin{equation}
\label{constant}
    \delta p+\rho\delta\phi_T+\Psi_{sh}=\mathrm{constant}\equiv h,
\end{equation}
where the constant $h$ needs to be obtained from the boundary conditions (which otherwise cannot be satisfied).\footnote{It cannot in general be set to zero, as done in Eq.~(31) of  \citet{2003ApJ...588..975C}, where it should in fact be an arbitrary function of $r$. However, this does not appear to affect the equations actually solved by those authors, presented in their Sects.~3.1.2 and 3.1.3.}

Choosing a spherical coordinate system $(r,\theta,\varphi)$ aligned with the axis of rotation, the centrifugal potential perturbation can be written as
\begin{eqnarray}
\label{centrifugal perturbation}
\delta\phi_{c}(r,\theta)&=&-\frac{1}{2}\Delta(\Omega^{2})r^{2}\sin^{2}\theta \nonumber  \\
&=&-\frac{1}{3}\Delta(\Omega^{2})r^{2}[P_{0}(\cos\theta)-P_{2}(\cos\theta)],
\end{eqnarray}
where $\Delta(\Omega^2)$ is the change in the squared angular frequency, and $P_{0}(\cos\theta)=1$ and $P_{2}(\cos\theta)=\frac{1}{2}(3\cos^{2}\theta-1)$ are the Legendre polynomials of order $\ell=0$ (monopole) and $\ell=2$ (quadrupole), respectively.

The incompressibility condition, $\Theta\equiv\vec\nabla\cdot\vec\xi=0$, forces the monopole component of the radial displacement field to vanish,\footnote{Here and below, we use the notation
\begin{equation}
    \label{notation}
    F(r,\theta)=\sum_{\ell}F_\ell(r)P_\ell(\cos\theta)
\end{equation}
for any function $F(r,\theta)$ expanded in terms of Legendre polynomials.} $\xi^r_0(r)=0$. Thus, motivated by the form of the centrifugal force, we assume %use the ansatz of 
a purely quadrupolar displacement, with
\begin{equation}
\label{radial quadrupole}
    \xi^{r}(r,\theta)=\xi^{r}_{2}(r)P_{2}(\cos\theta).
\end{equation}
Again applying the incompressibility condition, and using the Legendre equation, the polar component takes the form
\begin{equation}
\label{polar in terms of radial displacement}
\xi^{\theta}(r,\theta)=\frac{1}{6r^{2}}\frac{d}{dr}\left[r^2\xi^r_2(r)\right]\frac{d}{d\theta}\left[P_2(\cos\theta)\right].
\end{equation}

Equation~(\ref{elastoestatic change 3}) implies that
\begin{equation}
\label{curl}
\vec\nabla\times\vec{f}_{sh}
=2\mu\vec{\nabla}\times\vec{\nabla} \cdot\tens{\Sigma} 
=0, 
\end{equation} 
where the nonzero components of the strain tensor are
\begin{eqnarray}
\Sigma^{rr}&=&\partial_{r}\xi^{r}, \\
\Sigma^{\theta\theta}&=&\frac{1}{r^{2}}\left(\partial_{\theta}\xi^{\theta}+\frac{\xi^{r}}{r}\right), \\
\Sigma^{\phi\phi}&=&\frac{1}{r^{2}\sin^{2}\theta}\left(\cot\theta\xi^{\theta}+\frac{\xi^{r}}{r}\right), \\
\Sigma^{r\theta}=\Sigma^{\theta r}&=&\frac{1}{2}\left(\partial_{r}\xi^{\theta}+\frac{1}{r^{2}}\partial_{\theta}\xi^{r}\right).
\end{eqnarray}
For the displacement field given by Eqs.~(\ref{radial quadrupole}) and (\ref{polar in terms of radial displacement}), Eq. (\ref{curl}) becomes an ordinary differential equation of one independent variable,
\begin{equation}
\label{ODE}
r^{4}\frac{d^{4}\xi^{r}_{2}}{dr^4}+8r^{3}\frac{d^{3}\xi^{r}_{2}}{dr^3}-24r\frac{d\xi^{r}_{2}}{dr}+24\xi_{2}^{r}=0,
\end{equation}
whose solution we conveniently write as 
\begin{equation}
\label{solution displacement}
\xi^{r}(r,\theta)=-\frac{\Delta(\Omega^2)}{3\Omega_K^2}R_\star \tilde\xi^r_2 (\tilde r)P_{2}(\cos\theta),
\end{equation}
with the dimensionless radial displacement function 
\begin{equation}
    \label{polynomial}
\tilde\xi^{r}_2(\tilde r)=A_{1}\tilde r+A_{3}\tilde r^{3}+\frac{A_{-2}}{\tilde r^{2}}+\frac{A_{-4}}{\tilde r^{4}}
\end{equation}
(see also \citealt{2000ApJ...543..987F,2019PASA...36...36G}). Here, we have defined the dimensionless radial coordinate, $\tilde r\equiv r/R_\star$, as well as the Keplerian (or ``break-up'') frequency,
\begin{equation}
    \Omega_K\equiv\left(\frac{GM_\star}{R_\star^3}\right)^{1/2},
\end{equation}
for a star of radius $R_\star$ and mass $M_\star$. The coefficients $A_{1}$, $A_{3}$, $A_{-2}$ and $A_{-4}$ are dimensionless constants to be obtained from the boundary conditions. For later reference, we also write down the explicit expression for the shear potential,
\begin{eqnarray}
\Psi_{sh}(r,\theta)&=&\Psi_{sh,2}(r)P_{2}(\cos\theta) \\ \nonumber
&=&\mu\frac{\Delta(\Omega^2)}{3\Omega_K^2}\left(7A_{3}\tilde r^{2}+\frac{2A_{-2}}{\tilde r^{3}}\right)P_{2}(\cos\theta).
\end{eqnarray} 

\subsection{Fluid displacements in a rotating star}
\label{sec:fluid}

We show below that, although the force balance equation for the fluid is the same as that for the elastic solid but with $\mu=0$ (and thus $\vec{f}_{sh}=0$), in some cases the displacement field for the fluid cannot be obtained by taking the limit $\mu\to 0$ in the results for the solid. One formal indication of this is that, for $\mu=0$, Eq.~(\ref{curl}) is trivially satisfied and therefore does not imply Eq.~(\ref{ODE}) and its solution, Eq.~(\ref{polynomial}). 

On a conceptual level, we note that, in the elastic solid, neighboring matter elements cannot be displaced independently without causing large stresses, and therefore the vector field $\vec{\xi}(\vec{x})$ is well defined, with no ambiguity (except for a solid-body rotation around the symmetry axis). However, in a uniform, incompressible fluid, all matter elements are equivalent and interchangeable, so there is no well-defined correspondence of fluid elements in different equilibrium states, even if these states are very similar. Thus, any ``small'' displacement field satisfying the incompressibility condition $\vec{\nabla}\cdot\vec{\xi}=0$ and the appropriate boundary conditions is equally valid. This also implies that, although Eq.~(\ref{constant}) with $\Psi_{sh}=0$ can be used to determine the Eulerian pressure perturbation\footnote{Clearly $\delta p$ must in general be nonzero and nonspherical, contrary to statements in Sect.~3.1.1 of \citet{2003ApJ...588..975C}. However, these do not appear to affect the equations actually solved in that paper and thus its final results (C. Cutler and B. Link, private communications).} $\delta p(r,\theta)$, the ambiguity in the displacement field does not allow us to uniquely define a Lagrangian pressure perturbation, $\Delta p=\delta p+\vec{\xi}\cdot\vec{\nabla}p$. 

Of course, the assumption of uniform, incompressible matter is an extreme idealization that does not apply to a real neutron star, in which the strong gravity causes a density gradient, and variations in the centrifugal force (or other forms of forcing) change the density. Furthermore, the composition of neutron star matter is not uniform. In equilibrium, the relative abundances of particle species (neutrons, protons, electrons, and possibly other particles in the core; different nuclear species, electrons, and neutrons in the crust) are functions of the local pressure. However, if the pressure in a certain matter element is changed, its composition (here denoted by a generic label $Y$ that does not require a precise definition) does not respond instantaneously but on a finite timescale set by the required strong or weak interaction processes. Over short times (such as typical neutron star oscillation periods), the composition can be taken to be (approximately) conserved in each matter element, and the density must be treated as a function of (at least) two variables, namely pressure and composition, $\rho(p,Y)$ \citep{1987MNRAS.227..265F,1992ApJ...395..240R}, corresponding to a nonbarotropic (or "baroclinic") fluid. Over long timescales, such as the spin-down time of a pulsar, the composition in each fluid element of the neutron star core adjusts, staying close to its equilibrium value at the local pressure \citep{1995ApJ...442..749R}, so the matter can be regarded as barotropic, with the density as a unique function of pressure, $\rho(p)$. This may not be a good approximation for the neutron star crust, where successive equilibria are not a continuum, but they are separated by potential barriers \citep{2015MNRAS.453L..36G}.

In a nonbarotropic star with a changing centrifugal force, the displacement field $\vec\xi(\vec x)$ is much better defined than in the uniform, incompressible idealization. The force balance equation,
\begin{equation}
\label{fluid equilibrium}
-\vec{\nabla}p-\rho\vec{\nabla}\phi_T=0,
\end{equation}
implies that $\vec{\nabla}p$ and $\vec{\nabla}\phi_T$ are everywhere parallel to each other. Therefore, in a given equilibrium state, we can regard the pressure as a function of the total potential, $p=p(\phi_T)$. Similarly, the curl of Eq.~(\ref{fluid equilibrium}) implies that $\rho=\rho(\phi_T)$, and, from the equation of state, $\rho=\rho(p,Y)$, we conclude $Y=Y(\phi_T)$. Thus, in any given equilibrium state, there is a family of nested equipotential surfaces (spheroids) on each of which all four quantities ($\phi_T$, $p$, $\rho$, and $Y$) are constant, and, if $Y$ is conserved in each fluid element, it allows to identify the equipotentials containing the same fluid elements to be identified in different equilibrium states. Special cases of these equipotentials are the stellar surface (as long as the outermost layer is fluid) or any phase transition or density discontinuity between two fluid regions within the star. This also means that, for a small perturbation connecting two equilibrium states, the Lagrangian changes $\Delta p$ and $\Delta\rho$ must take the same value everywhere on a given equipotential (and of course $\Delta Y=0$). Since the Eulerian changes are well defined, we can use any of the fluid variables to constrain the displacement field (e.g., $\vec{\xi}\cdot\vec{\nabla}p=\Delta p-\delta p$); therefore, the component of the displacement field perpendicular to the equipotentials is uniquely fixed, and the parallel component can be obtained from solving $\vec{\nabla}\cdot\vec{\xi}=-\Delta\rho/\rho$ with the appropriate symmetry and boundary conditions (as it was done to obtain Eq.~(\ref{polar in terms of radial displacement}) from Eq.~(\ref{radial quadrupole})).

The equipotentials still exist (with $p$, $\rho$, and $\phi_T$ constant on each of them) in a barotropic fluid (likely a good approximation for the core of a neutron star that spins down slowly) and in the idealized uniform and incompressible case. However, in these cases, there is no guarantee that the set of fluid elements lying on one equipotential in some equilibrium state still lie on one equipotential in another equilibrium, as they can be interchanged (allowing them to adjust to the local pressure) without changing the structure of the equilibrium state. Still, we can choose to identify fluid elements on equipotentials in two different equilibria, using a conserved integral quantity, such as the total baryon number enclosed by each equipotential, to match equipotentials in different equilibrium states. Although this choice is arbitrary (analogous to fixing the gauge for the electromagnetic potentials), it has the advantage of including the stellar surface and internal discontinuities as special cases of such equipotentials (if the adjacent matter is fluid). 

\begin{figure}
    \centering
    \resizebox{\hsize}{!}{\includegraphics{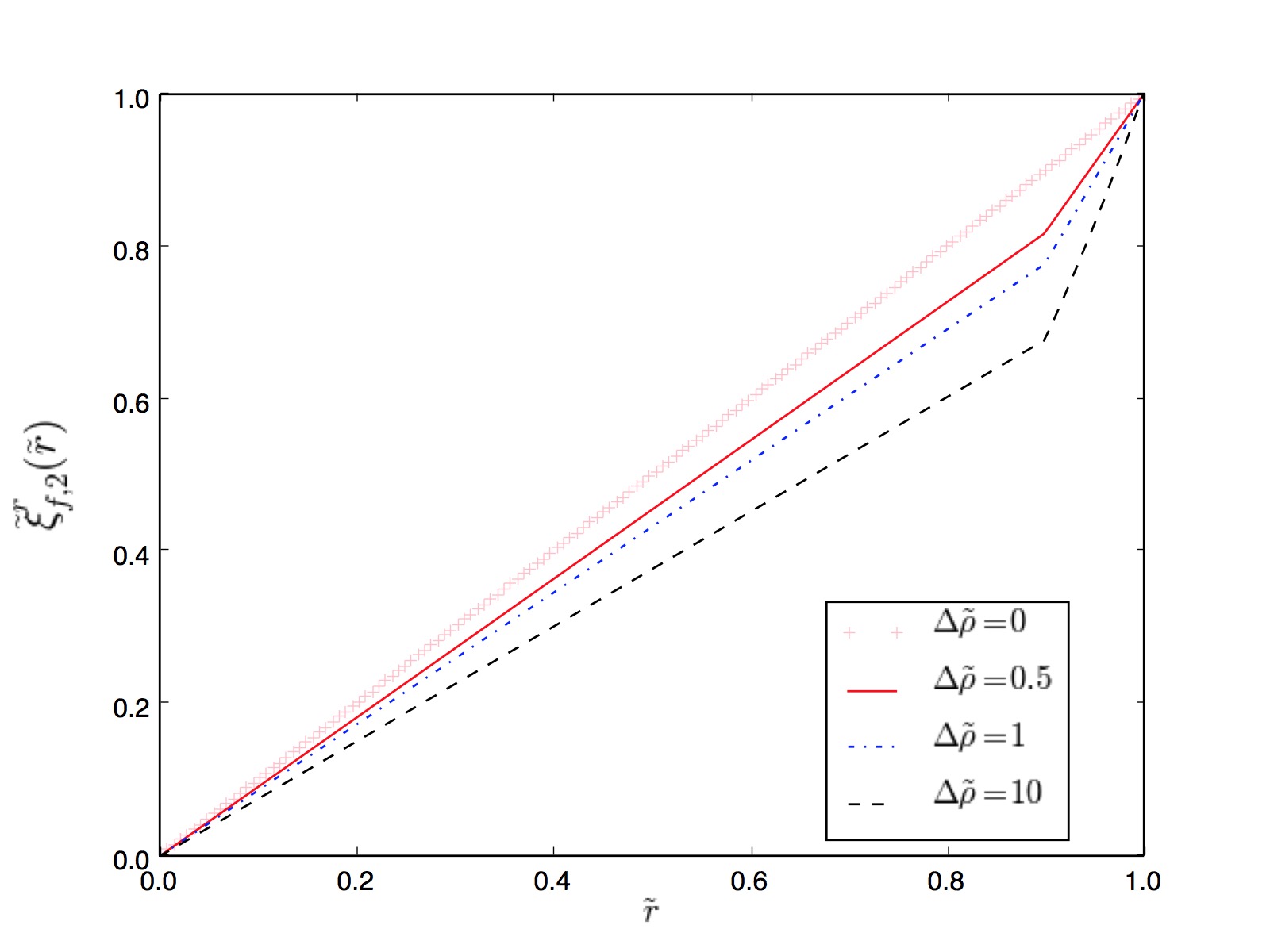}}
    \caption{Radial dependence of the normalized quadrupolar radial displacement field, $\tilde\xi^r_{f,2}(\tilde r)$, of the equipotential surfaces in a completely fluid star with uniform densities, $\rho_s$ in the crust ($\tilde r_i<\tilde r<1$) and $\rho_f=\rho_s(1+\Delta\tilde\rho)$ in the core ($0<\tilde r<\tilde r_i$), with a dimensionless interface radius $\tilde r_i=0.9$ and different values of $\Delta\tilde\rho$.}
    \label{fig:radial displacement fluid}
\end{figure}

Thus, in our incompressible, uniform density model, we proceed by writing Eq.~(\ref{constant}) with $\Psi_{sh}=0$ as
\begin{equation}
\label{fluid-constant}
    \Delta p-\vec{\xi}\cdot\vec{\nabla}p+\rho\delta\phi_T=h.
\end{equation}
In what follows, we assume a slow rotation, $\Omega\ll\Omega_K$, so the background equilibrium model can be taken as spherically symmetric, with $p=p(r)$ satisfying the hydrostatic equilibrium condition $dp/dr=-\rho g(r)$, where $g(r)=|\vec{\nabla}\phi_g|$ is the local gravity, and with $\Delta p=\Delta p(r)$. Since $\xi^r(r,\theta)$ is purely quadrupolar, the monopole component of Eq.~(\ref{fluid-constant}) gives
\begin{equation}
\label{fluid pressure}
    \Delta p_0(r)=-\rho\delta\phi_{T0}(r)+h,
\end{equation}
whereas the quadrupole part yields
\begin{equation}
\label{fluid displacement}
    \xi^r_2(r)=-\frac{\delta\phi_{T2}(r)}{g(r)}\approx-\frac{\Delta(\Omega^2)r^2}{3g(r)}.
\end{equation}
In the last equality, we also used the Cowling approximation, namely neglecting the gravitational potential perturbation, $\delta\phi_g$, so $\delta\phi_{T}\approx\delta\phi_{c}$. Equation~(\ref{fluid displacement}) determines the shape of all equipotentials in the fluid regions of the star, particularly the surface and the crust-core interface, as long as both the crust and the core can be modeled as fluids. For our toy model, with a uniform crust and a uniform core, Fig.~\ref{fig:radial displacement fluid} shows the radial dependence of this displacement field. The change in the slope at the interface $\tilde r=\tilde r_i$ is caused by the assumed density discontinuity, quantified by $\Delta\tilde\rho\equiv(\rho_f-\rho_s)/\rho_s$, through its effect on $g(r)$. We note that, in the crustal region, $\tilde r_i<\tilde r<1$, Eq.~(\ref{fluid displacement}) implies
\begin{equation}
\label{normalized fluid displacement}
    \tilde\xi_2^r(\tilde r)=\frac{(\tilde r_i^3\Delta\tilde\rho+1)\tilde r^4}{\tilde r_i^3\Delta\tilde\rho+\tilde r^3},
\end{equation}
which does not have the polynomial form of the elastic displacement (Eq.~(\ref{polynomial})), so these two functions can at best agree at a few discrete values of $\tilde r$.

\subsection{Boundary conditions}
\label{sec:boundary}

At the interfaces between domains with different properties (in the present case, the core and crust of the neutron star), we must impose continuity of the perpendicular displacement field, 
$\vec{n}\cdot\vec{\xi}$, and of the stress across the boundary, $\vec{n}\cdot\tens{T}$, where $\vec{n}$ is the normal vector of the interface. At external boundaries (the stellar surface, separating the crust from the exterior, approximated as a vacuum), the first of these conditions becomes meaningless, and the second implies $\vec{n}\cdot\tens{T}=0$ at the boundary. 

If both the core and the crust of the star are uniform, incompressible fluids (of possibly different densities), the displacement of all equipotential surfaces (including the interface and the surface) is given by Eq.~(\ref{fluid displacement}). Therefore the continuity condition on $\vec{n}\cdot\vec{\xi}$ is trivially satisfied, and the boundary conditions on $\vec{n}\cdot\tens{T}$ are only required for the calculation of the pressure perturbations. (In Eq.~(\ref{fluid pressure}), the constant $h$ could be different in the two domains.)

If, on the other hand, the crust supports elastic stresses, the displacement field does not follow equipotential surfaces and thus does not satisfy Eq.~(\ref{fluid displacement}). This applies, in particular, to the displacement of the crust-core interface, where the fluid displacement field cannot be given by this equation and thus the continuity condition on $\vec{n}\cdot\vec{\xi}$ becomes inapplicable. 

On the other hand, we do need to apply the boundary conditions on $\vec{n}\cdot\tens{T}$. The displacement of the stellar surface clearly follows the matter displacement field evaluated at the stellar radius, so the condition of zero external stress implies a condition on the Lagrangian perturbation of the stress tensor
\begin{equation}
[\Delta\tens{T}_s\cdot\vec{n}]_{r=R_{\star}}=0.
\label{boundary condition lagrangian stress surface}
\end{equation}

At the crust-core interface, the condition is less obvious. As the pressure at the interface increases, strong interaction processes quickly dissolve nuclei or ``pasta'' structures \citep{1983PhRvL..50.2066R} in the adjacent crust into free nucleons and electrons (and the opposite process if the pressure decreases), converting some solid to fluid (or vice versa). We follow the previous literature (e.g., \citealt{1971AnPhy..66..816B,2000ApJ...543..987F,2019PASA...36...36G}) in ignoring this effect, which affects only a small fraction $\sim\Delta(\Omega^2)/\Omega_K^2$ of the matter, and therefore we also assume that the interface moves with the local displacement field of the solid, yielding
\begin{equation}
[\Delta\tens{T}_s\cdot\vec{n}]_{r=R_{i}}=
[\Delta\tens{T}_f\cdot\vec{n}]_{r=R_{i}}
\label{boundary condition lagrangian stress}
\end{equation}
at the interface radius $R_i$ between the fluid core (labeled by $f$) and the solid crust (labeled by $s$).\footnote{Since $\Delta T^{rr}=\delta T^{rr}-\rho g\xi^r$, these conditions are equivalent to imposing continuity of the Eulerian stress perturbation (as done in \citealt{2000MNRAS.319..902U} and \citealt{2003ApJ...588..975C}) if the density is continuous across the respective boundary, but not if there is a density discontinuity, as in the general case considered here.}

In our model, these boundary conditions become
\begin{eqnarray}
\label{traction rth interface}
\Sigma^{r\theta}|_{R_{i}}=\Sigma^{r\theta}|_{R_{\star}}&=&0,\\
\label{boundary condition interface}
(\Delta p_{s}-2\mu\Sigma^{rr})|_{R_{i}}&=&\Delta p_{f}|_{R_{i}},\\
\label{boundary condition surface}
(\Delta p_{s}-2\mu\Sigma^{rr})|_{R_{\star}}&=&0.
\end{eqnarray}
Equations~(\ref{traction rth interface}) can be rewritten as
\begin{equation}
\label{no shear}
    \left[r^2\frac{d^2\xi^r_2}{dr^2}+2r\frac{d\xi^r_2}{dr}+4\xi^r_2\right]_{r=R_i,R_{\star}}=0,
\end{equation}
providing two independent linear homogeneous equations relating the four constants $A_k$ in Eq.~(\ref{polynomial}), 
\begin{equation}
    \label{homogeneous}
    \left[3A_1\tilde r+8A_3\tilde r^3+3A_{-2}\tilde r^{-2}+8A_{-4}\tilde r^{-4}\right]_{ r= R_i,R_{\star}}=0.
\end{equation}
Equation~(\ref{boundary condition interface}) is most conveniently written as
\begin{equation}
    \label{interface}
    \left[(\rho_f-\rho_s)(g\xi^r+\delta\phi_T)-\Psi_{sh}-2\mu\partial_r\xi^r\right]_{r=R_i}=h_f-h_s.
\end{equation}
This equation has a monopolar component, yielding $h_f-h_s=(\rho_f-\rho_s)\delta\phi_{T,0}(R_i)\approx-(1/3)(\rho_f-\rho_s)\Delta(\Omega^2)R_i^2$, and a quadrupole component,
\begin{equation}
    \label{interface quadrupole}
    (\rho_f-\rho_s)\left[g\xi^r_2+\delta\phi_{T,2}\right]_{r=R_i}=\left[\Psi_{sh,2}+2\mu\frac{d\xi^r_2}{dr}\right]_{r=R_i},
\end{equation}
or, equivalently,
\begin{eqnarray}
    \label{interface condition}
    (\rho_f-\rho_s)R_\star\left[-g(R_i)\left(A_1\tilde r_i+A_3\tilde r_i^3+\frac{A_{-2}}{\tilde r_i^2}+\frac{A_{-4}}{\tilde r_i^4}\right)+g(R_\star)\tilde r_i^2\right] \nonumber \\
    =\mu\left(-2A_1+A_3\tilde r_i^2+\frac{6A_{-2}}{\tilde r_i^3}+\frac{8A_{-4}}{\tilde r_i^5}\right).
\end{eqnarray}
We note that in Eqs.~(\ref{interface quadrupole}) and (\ref{interface condition}) the left-hand side is proportional to $\rho_f-\rho_s$ and the right-hand side is proportional to $\mu$. Thus, in the limit $\mu\to 0$ with $\rho_f\neq\rho_s$, Eq.~(\ref{interface quadrupole}) becomes $\xi^r_2(R_i)=-\delta\phi_{T,2}(R_i)/g(R_i)$, that is, the interface becomes an equipotential surface (see Eq.~(\ref{fluid displacement})), because the solid behaves essentially like a fluid. On the other hand, for $\rho_f-\rho_s\to 0$ with $\mu\neq 0$, the boundary condition is very different, namely $[\Psi_2+\mu\partial_r\xi^r_2]_{r=R_i}=0$, a homogeneous linear condition involving up to third derivatives of $\xi^r_2(r)$. Thus, as further discussed in Sect.~\ref{sec:two domains}, the two limits $\Delta\tilde\rho\to 0$ and $\mu\to 0$ do not commute, as they yield qualitatively different interface conditions. 

Similarly, at the surface of the star, Eq.~(\ref{boundary condition surface}) yields
\begin{equation}
    \label{surface}
    \left[\rho_s(g\xi^r+\delta\phi_T)+\Psi_{sh}+2\mu\partial_r\xi^r\right]_{r=R_\star}=h_s,
\end{equation}
implying $h_s=\rho_s\delta\phi_{T,0}(R_\star)\approx-(1/3)\rho_s\Delta(\Omega^2)R_\star^2$ for the monopolar component and 
\begin{equation}
    \label{surface quadrupole}
    \rho_s\left[g\xi^r_2+\delta\phi_{T,2}\right]_{r=R_\star}=-\left[\Psi_{sh,2}+2\mu\frac{d\xi^r_2}{dr}\right]_{r=R_\star},
\end{equation}
or
\begin{eqnarray}
    \label{surface condition}
    \rho_s R_\star g(R_\star)\left[-(A_1+A_3+A_{-2}
    +A_{-4})+1\right] \nonumber \\
    =-\mu\left(-2A_1+A_3+6A_{-2}+8A_{-4}\right)
\end{eqnarray}
for the quadrupolar component. 
In the limit $\mu\to 0$, the right-hand side of Eqs.~(\ref{surface quadrupole}) and (\ref{surface condition}) vanishes, so the stellar surface also becomes an equipotential. 

In all cases, the four linear Eqs.~(\ref{homogeneous}), (\ref{interface condition}), and (\ref{surface condition}) yield a unique solution for the four constants $A_k$, thus fully determining the displacement field in the neutron star crust.

\subsection{Displacement field for a uniform star}
\label{sec:one domain}

\begin{figure}
    \centering
    \resizebox{\hsize}{!}{\includegraphics[width=0.55\textwidth]{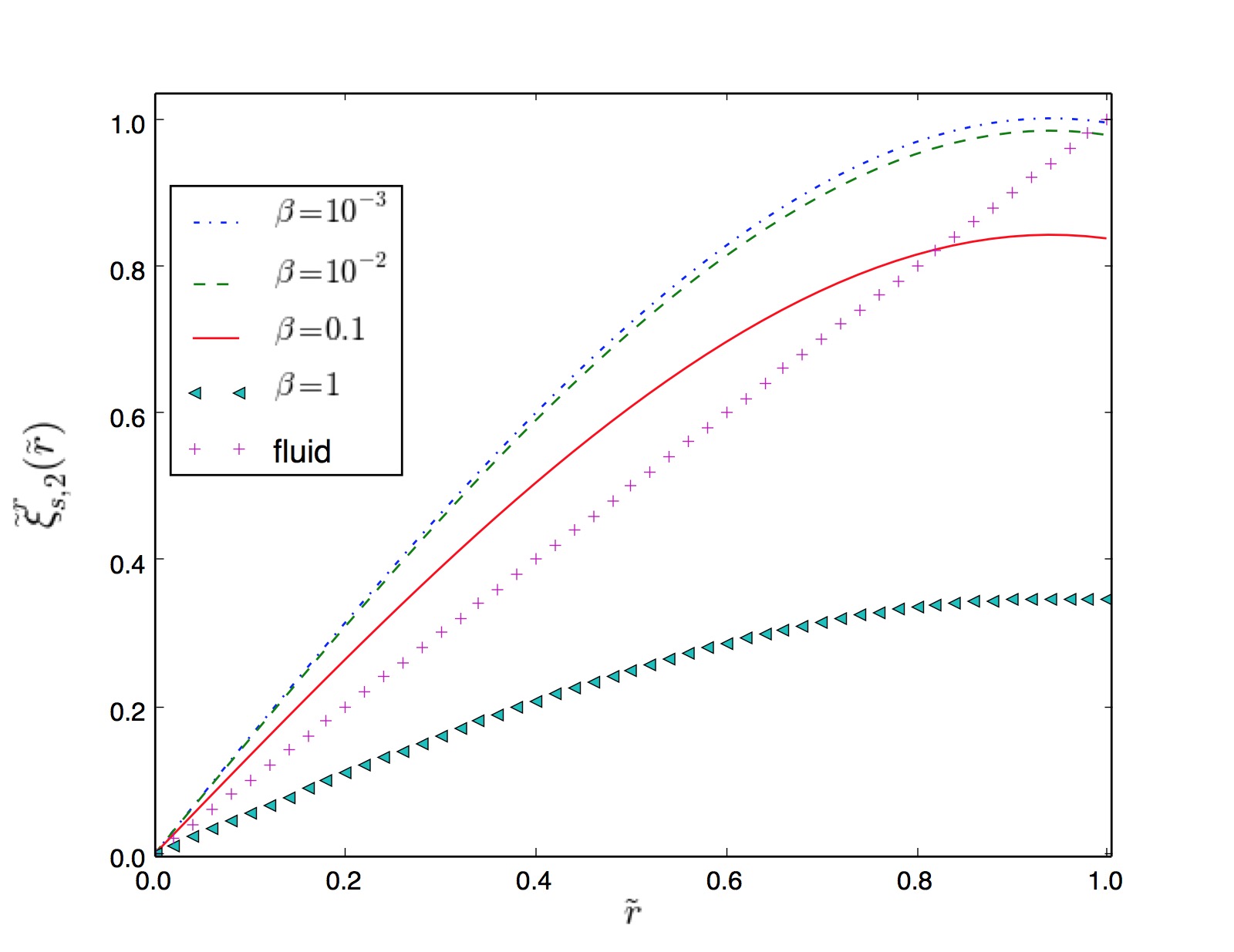}}
    \caption{Dimensionless radial displacement $\tilde\xi^r_2(\tilde r)$ 
    in the uniform solid star for different values of the parameter $\beta$ defined in Eq.~(\ref{beta}), compared to the case of an uniform fluid, with the displacement field in the latter case defined in terms of equipotentials, as explained in Sect.~\ref{sec:fluid}.}
    \label{fig:radial displacement B&P}
\end{figure}

Before presenting the full analytic solution for the displacement field in a two-domain star (crust and core with different properties), it is useful to consider the simple cases of a completely uniform (single-domain) star, both as a fluid and as a uniform elastic solid. For the  two-domain model, %former, 
with our convention of assuming that fluid elements move with the equipotentials (which they would, in the case of a nonbarotropic star), the dimensionless displacement field can be obtained from Eq.~(\ref{fluid displacement}) as
\begin{equation}
\label{uniform fluid}
    \tilde\xi^r_{f,2}=\tilde r,
\end{equation}
where $\tilde r\equiv r/R_\star$ is the normalized radius. For the one-domain model (the case considered by \citealt{1971AnPhy..66..816B}), regularity of Eq.~(\ref{polynomial}) at $r=0$ implies $A_{-2}=A_{-4}=0$, which, together with Eq.~(\ref{no shear}) (imposed only at the surface) and Eq.~(\ref{surface condition}), gives 
\begin{equation}
\label{uniform solid}
    \tilde\xi^r_{s,2}=\frac{\frac{8}{5}\tilde r-\frac{3}{5}\tilde r^3}{1+\frac{19}{5}\beta},
\end{equation}
with the dimensionless parameter\footnote{We note that $\beta=(c_t/v_K)^2$, where $c_t=\sqrt{\mu/\rho_s}\sim 10^8\mathrm{cm\,s}^{-1}$ is the (roughly constant) speed of transverse shear waves in the elastic crust and $v_K=\sqrt{g(R_\star)R_\star}=\sqrt{GM_\star/R_\star}\sim 10^{10}\mathrm{cm\,s}^{-1}$ is the Keplerian rotation velocity at the stellar surface. For a star of uniform density $\rho_s$, this expression is equivalent to $\beta=\mu/(2p_c)$, where $p_c$ is the pressure at the center of the star.}
\begin{equation}
\label{beta}
    \beta\equiv\frac{\mu}{\rho_s g(R_\star)R_\star}=10^{-4}\frac{\mu_{30}}{\rho_{s,14}g_{14}R_{6}},
\end{equation}
where the numerical subscripts stand for the exponents in the reference magnitudes for neutron stars, expressed in cgs units, e.g., $\mu_{30}\equiv\mu/(10^{30}\,\mathrm{erg\,cm^{-3}})$. Figure ~\ref{fig:radial displacement B&P} compares these two cases, considering several values of $\beta$, which are chosen unrealistically large for the purpose of illustration.

The radial displacement at the surface ($\tilde r=1$) of the elastic star is smaller by only a fractional amount ${\cal O}(\beta)$ than that of its fluid counterpart. On the other hand, the internal displacement is very different, as it increases linearly with $r$ in the fluid case, whereas in the solid case it has a convex shape with a maximum at $r/R_\star=(8/9)^{1/2}\approx 0.94$, where the displacement is slightly larger (by a factor of $32\sqrt{2}/45\approx 1.006$) than on the surface. As discussed in Sect.\ref{sec:fluid}, the equilibrium displacement field within a uniform or barotropic fluid is a matter of convention. Our choice is based on the case of a nonbarotropic fluid, in which equipotential surfaces of one equilibrium must deform into equipotential surfaces of another.  

If instead the matter is a nonbarotropic, elastic solid, there is a conflict between the tendency to follow equipotentials (as in Eq.~(\ref{uniform fluid})) and the tendency to minimize the shear stresses in the crust (as in Eq.~(\ref{uniform solid})).

\subsection{Displacement field for two uniform domains}
\label{sec:two domains}

The uniform-density model of \citet{2000ApJ...543..987F} accounts for the shear stresses, but of course it does not include the buoyancy force associated with the density gradient present in a realistic neutron star. This issue is addressed in the present paper (as in \citealt{2019PASA...36...36G}) by allowing for a density difference between the core and the crust, which makes the crust float on top of the core, so their interface would be an equipotential if shear forces could be ignored. This is a strong simplification, which reduces a gradient in density and composition (from the center to the surface of the star) to a single step. However, it has the advantage of being analytically calculable and introducing a single additional parameter, $\Delta\tilde\rho\equiv(\rho_f-\rho_s)/\rho_s$, thus making it easier to analyze than the more realistic numerical calculations of \citet{2003ApJ...588..975C} or \citet{2020MNRAS.491.1064G}. Using the average densities in the core and crust for the range of
neutron star models shown in Fig.~1 of \citet{2018arXiv180404952F}, one obtains a density contrast in the range $10\lesssim\Delta\tilde\rho\lesssim 30$. 

The boundary condition at the core-crust interface (Eqs.~[\ref{interface quadrupole}] and [\ref{interface condition}]) implies that our model reduces to the uniform-density model of \citet{2000ApJ...543..987F} when $\Delta\tilde\rho\ll\beta$ (except that we use the Cowling approximation, $\delta\phi_g=0$, whereas \citealt{2000ApJ...543..987F} calculate $\delta\phi_g[r,\theta]$ self-consistently). On the other hand, considering also the surface boundary condition (Eqs.~[\ref{surface quadrupole}] and [\ref{surface condition}]), one can see that our model must approach the pure fluid case (with the displacement field of the crust-core interface and stellar surface following equipotentials) when $\beta\ll\min\{\Delta\tilde\rho, 1\}$.

The full solutions for the coefficients of the displacement field in the elastic crust with the boundary conditions presented in Sect.~\ref{sec:boundary} are given in Appendix A, and specific cases are plotted in Figs.~\ref{fig:crust} and \ref{fig:rdispsurfint2}.

Evaluating the full solution at the core-crust interface yields
\begin{equation}
\label{radial displacement interface solid}
\tilde\xi_{2}^{r}(\tilde r_{i})=\frac{\tilde{r}_{i}(3d_{\Delta1}\tilde{t}\Delta\tilde{\rho} +11d_{\Delta2}\Delta\tilde{\rho}\beta+11n_{0}\beta)}{3w_{0}d_{\Delta1}\tilde{t}\Delta\tilde{\rho}+11w_{0}d_{\Delta2}\Delta\tilde{\rho}\beta+11d_{1}\beta+24d_{2}\tilde{t}\beta^{2}},
\end{equation}
and at the stellar surface,
\begin{equation}
\label{radial displacement surface solid}
\tilde\xi_{2}^{r}(1)=\frac{3w_{0}d_{\Delta1}\tilde{t}\Delta\tilde{\rho}+11\tilde{r}_i^5n_{0}\Delta\tilde{\rho}\beta+11d_{1}\beta}{3w_{0}d_{\Delta1}\tilde{t}\Delta\tilde{\rho}+11w_{0}d_{\Delta2} \Delta\tilde{\rho}\beta+11d_{1}\beta+24d_{2}\tilde{t}\beta^{2}},
\end{equation}
where the coefficients $n_k, d_k, d_{\Delta k}$ (for various integers $k$), defined in Appendix A,  depend only on the dimensionless crust thickness $\tilde t\equiv 1-\tilde r_i$, whereas $w_0=(\Delta\tilde{\rho}+1)/(\tilde{r}_{i}^{3}\Delta\tilde{\rho}+1)$. All these coefficients have been defined in such a way that they approach 1 as $\tilde t\to 0$. 

In these expressions, it is clear that the limits $\tilde t\to 0$, $\Delta\tilde\rho\to 0$, and $\beta\to 0$ do not commute. Here we discuss them one by one. 

In the simplest case $\tilde t=0$, in which the crust disappears, so $\beta$ and $\Delta\tilde\rho$ become irrelevant and $\tilde\xi_2^r(\tilde r_i)=\tilde\xi_2^r(1)=1$, since both quantities correspond to the surface of a purely fluid star of uniform density. 

In the uniform-density case ($\Delta\tilde\rho=0$), the exact solutions are
\begin{equation}
\label{radial displacement interface app FLE}
    \tilde\xi^r_2(\tilde r_i)=\frac{\tilde{r}_in_0/d_{1}}{1+\frac{24}{11}\frac{d_2}{d_{1}}\beta \tilde{t}}
\end{equation}
and
\begin{equation}
\label{radial displacement surface app FLE}
    \tilde\xi^r_2(1)=\frac{1}{1+\frac{24}{11}\frac{d_2}{d_{1}}\beta \tilde{t}}.
\end{equation}
In this case, it can be seen that the displacement field $\tilde\xi^r_2(r)$ depends on the shear modulus only through the constant factor of $1+(24/11)(d_2/d_{1})\beta\tilde{t}$ in the denominator, so its value affects only the amplitude but not the shape of the displacement field. Taking the limit $\beta\to 0$, this solution, like the completely solid star of \citet{1971AnPhy..66..816B}, approaches the fluid limit at the surface but not in the interior (particularly at the interface), where the fluid displacement field was chosen to follow equipotentials. This limit is illustrated by the upper panel of Fig.~\ref{fig:crust}, where $\Delta\tilde\rho=0\ll\beta=10^{-4}\ll 1$.

In the arguably more realistic case of $\beta\ll\min\{\Delta\tilde\rho,1\}$, we can expand %for small $\beta$ around zero 
the normalized radial displacement at the interface as 
\begin{equation}
\tilde\xi_{2}^{r}(\tilde r_{i})=\frac{\tilde r_{i}}{w_0}\left[1+\frac{11(w_0n_0-d_1)}{3w_0 d_{\Delta 1}{\Delta\tilde{\rho}}\tilde t}\,\beta+{\cal O}(\beta^2)\right],
\end{equation}
and at the surface as
\begin{equation}
\tilde\xi_{2}^{r}(1)=1-\frac{11(w_0d_{\Delta 2}-\tilde r_i^5n_0)}{3w_0d_{\Delta 1}\tilde{t}}\,\beta+{\cal O}(\beta^2),
\end{equation}
where we see that the term independent of $\beta$ corresponds to the case of two fluids with different densities (Eq.~(\ref{fluid displacement})), and the crust elasticity comes in as a small correction $\propto\beta$.
However, we can see in the lower panel of Fig.~\ref{fig:crust} that within the crust ($\tilde r_i<\tilde r<1$), the formal solution in the limit $\beta\to 0$ does not correspond exactly to our fluid displacement field (chosen %arbitrarily 
to follow equipotentials).

\begin{figure}
\centering
\resizebox{\hsize}{!}{\includegraphics{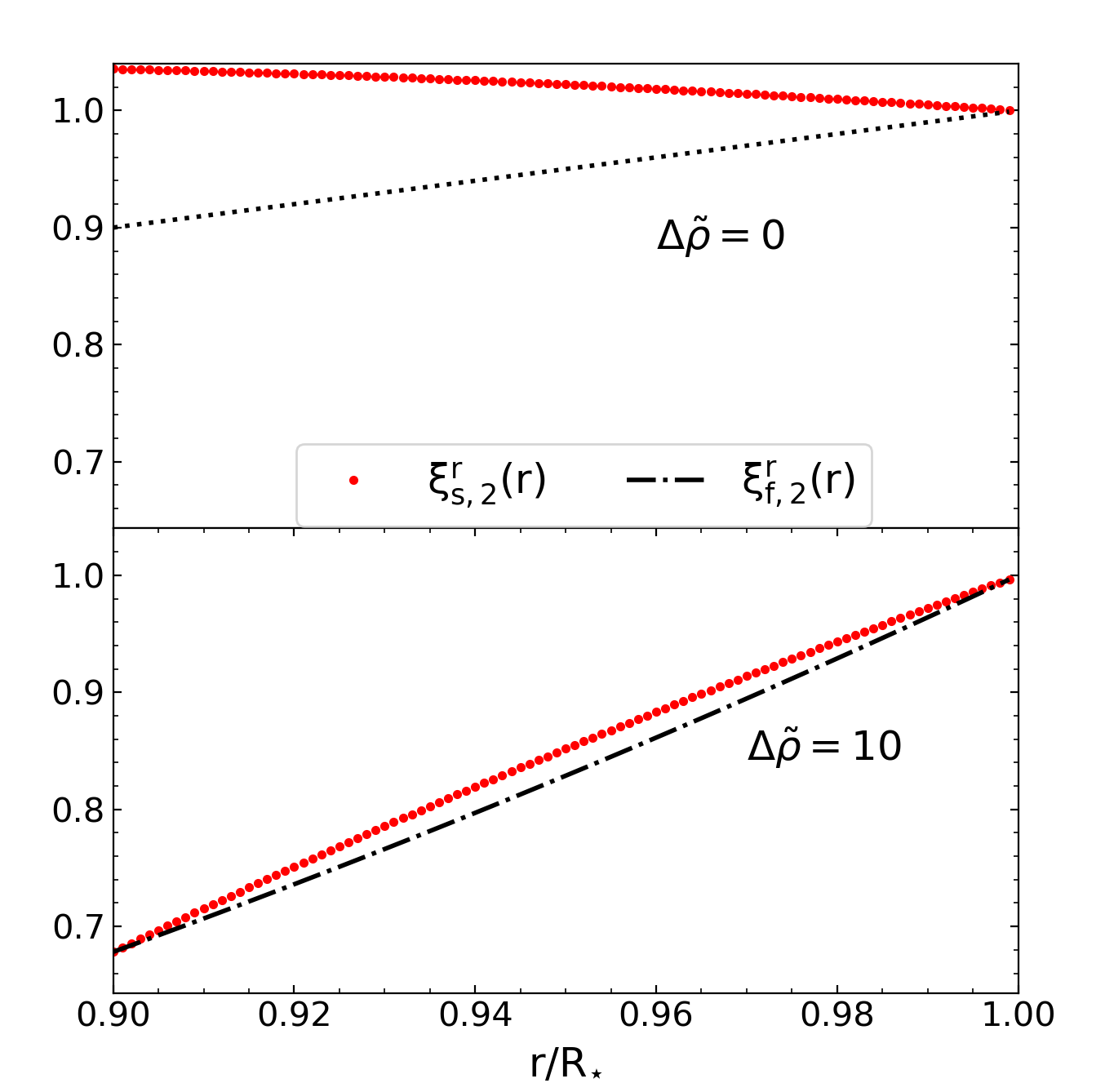}}
\caption{Normalized radial displacements $\tilde\xi^{r}_{2}(\tilde r)$ in the neutron star crust with interface radius $\tilde r_i=0.9$, both for purely fluid matter (dash-dotted line) and for an elastic crust with $\beta=10^{-4}$ (red dots). The upper panel corresponds to a very small density difference, $\Delta\tilde{\rho}\ll \beta$, whereas the lower panel shows a star with a much denser core, $\Delta\tilde{\rho}=10$.}
\label{fig:crust}
\end{figure}

\begin{figure}
\centering
\resizebox{\hsize}{!}{\includegraphics[scale=0.48]{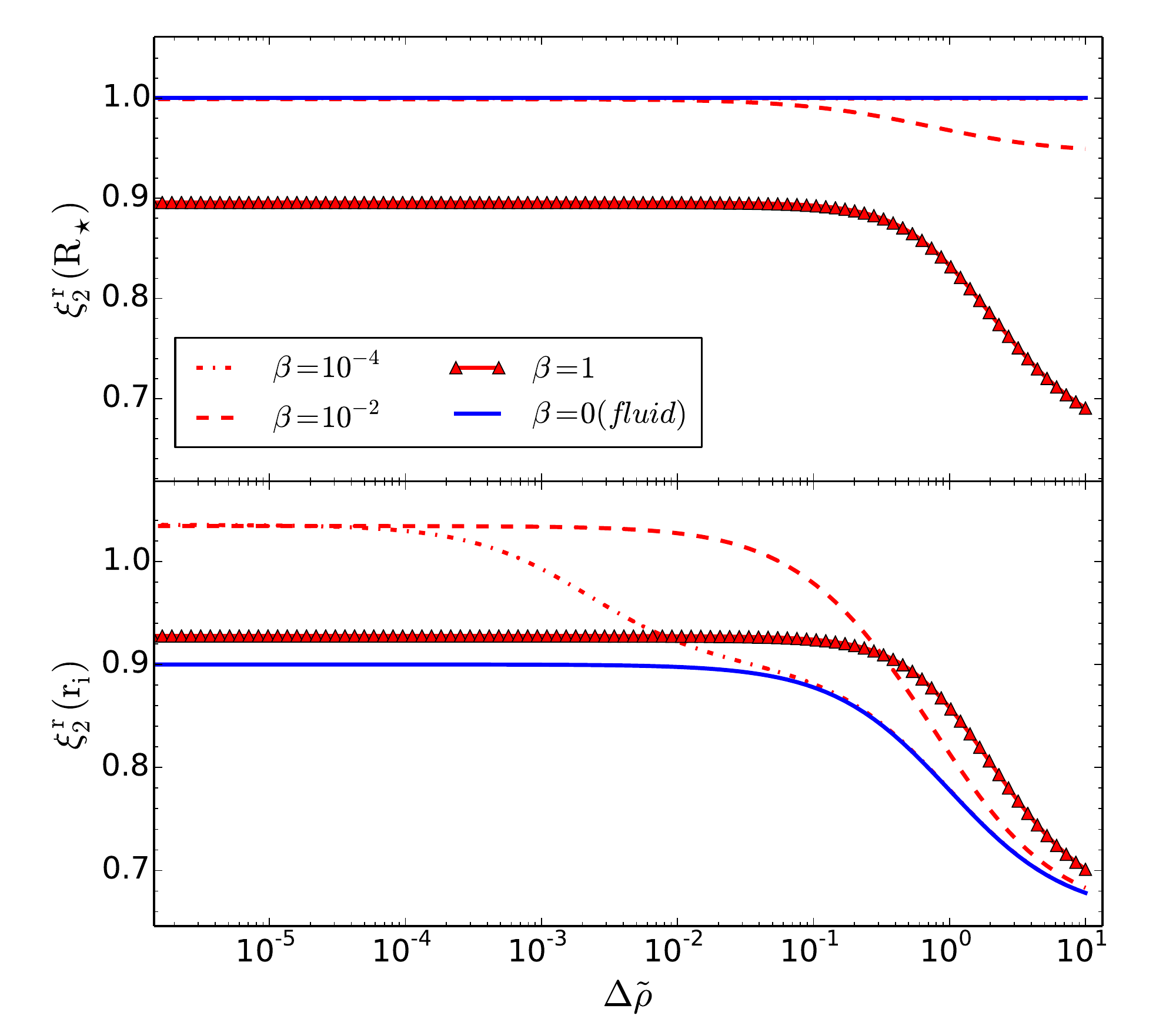}}
\caption{Normalized radial displacement $\tilde\xi^{r}_{2}$ at the stellar surface ($r=R_{\star}$; upper panel) and at the crust-core interface ($r=R_i=0.9 R_{\star}$; lower panel) as a function of the normalized density difference $\Delta\tilde\rho$ between the core and the crust. The blue line corresponds to a completely fluid star, whereas the other lines correspond to stars with an elastic crust, for different values of $\beta$.}
\label{fig:rdispsurfint2}
\end{figure}

\section{Modeling a glitch as a starquake}\label{sec:glitch}

\subsection{Rigidity parameter} \label{sec:rigidity}

In the starquake model, illustrated in Fig.~\ref{fig:esquemastarquake.pdf}, the change in the rotation rate in a glitch, $\Delta\Omega_{bc}$, is due to conservation of angular momentum as the star's moment of inertia $I$ changes by $\Delta I_{bc}$ when its crust breaks and relaxes,
\begin{equation}
\label{glitch conservation}
    \frac{\Delta\Omega_{bc}}{\Omega}=-\frac{\Delta I_{bc}}{I}.
\end{equation}
The change in the moment of inertia due to the starquake can be calculated as
\begin{equation}
\label{MoI relaxation}
    \Delta I_{bc}=\Delta I_{ac}-\Delta I_{ab},
\end{equation}
where the change from the initial, relaxed state "a" to the stressed state "b" is resisted by the crustal shear stresses, whereas the change from "a" to the relaxed state "c" can be modeled in terms of a purely fluid star (with $\mu=0$). Both changes $\Delta I_{ab}$ and $\Delta I_{ac}$ can be written in terms of the change in the squared rotation rate between starquakes, $\Delta(\Omega^2)$, and the induced normalized displacement fields as
\begin{eqnarray}
\label{change moment of inertia total}
\frac{\Delta I}{I}=\frac{\Delta(\Omega^2)}{3\Omega_K^2}\frac{\tilde\xi^{r}_{2}(R_{\star})+\Delta\tilde{\rho}\tilde r_{i}^{4}\tilde\xi^{r}_{2}(R_{i})}{1+\tilde r_{i}^{5}\Delta\tilde{\rho}}\equiv\frac{\Delta(\Omega^2)}{3\Omega_K^2}\Delta\tilde  I(\Delta\tilde\rho,\tilde t,\beta),
\end{eqnarray}
where, for the stress-free (``fluid'') displacement ($\Delta I_{ac}$), the dimensionless function $\Delta\tilde I$ must be evaluated at $\beta=0$. Using Eqs.~(\ref{MoI relaxation}) and (\ref{change moment of inertia total}) and assuming slow rotation and small $\beta$, the ``rigidity parameter'', defined by \citet{2003ApJ...588..975C} as the ratio between the change in moment of inertia at a starquake and between starquakes, can be approximated as
\begin{equation}
\label{rigidity general}
b \equiv -\frac{\Delta I_{bc}}{\Delta I_{ac}}\approx \frac{\Delta\tilde  I(\Delta\tilde\rho,\tilde t,0)-\Delta\tilde I(\Delta\tilde\rho,\tilde t,\beta)}{\Delta\tilde I(\Delta\tilde\rho,\tilde t,0)},%-\left( \frac{\Delta\Omega_{bc}^2}{\Delta\Omega_{ac}^2}\right).
\end{equation}
where we have neglected a term %Using equations~(\ref{glitch conservation}), (\ref{MoI relaxation}), and (\ref{change moment of inertia total}), one obtains 
%\begin{equation}
%\label{ratio of Omegas}
%\left|\frac{
$\Delta\Omega_{bc}/\Delta\Omega_{ac}\sim b(\Omega/\Omega_K)^2$, which is small in the limit of slow rotation,
%}\,\Delta\tilde I(\Delta\tilde\rho,\tilde t,0).
%\end{equation}
%Thus, for 
$\Omega\ll\Omega_K$.
%, this term can be neglected in equation~(\ref{rigidity general}). In the general case, we 
We can evaluate
\begin{eqnarray}
\label{function f general}
 \lefteqn{\Delta\tilde  I(\Delta\tilde\rho,\tilde t,\beta)=} \nonumber\\
&&[3d_{\Delta1}(w_{0}+\tilde r_i^5 \Delta\tilde\rho)\tilde{t}\Delta\tilde{\rho}+11\tilde{r}_i^5(2n_{0}+d_{\Delta2}\Delta\tilde\rho)\beta\Delta\tilde{\rho}+11d_{1}\beta] \nonumber\\
&&[3d_{\Delta1}w_{0}\Delta\tilde{\rho}\tilde{t}+11d_{\Delta2}w_{0} \Delta\tilde{\rho}\beta+11d_{1}\beta+24d_{2}\tilde{t}\beta^{2}]^{-1} \nonumber\\
&&(1+\tilde r_i^5 \Delta\tilde\rho)^{-1},
\end{eqnarray}
%which is the change between the states "a" and "b" due to the change in the squared rotation rate, $\delta(\Omega^2)_{ba}$, between those two states.
%because of the spin-down. 
%So, both the deformation of the core and crust contribute to the change in the MoI. If we use the boundary conditions of equation~(\ref{boundary condition lagrangian stress surface}), we get  For 
which, in the case without shear stresses in the crust (the purely fluid case; Eq.~(\ref{normalized fluid displacement})) becomes %yields
\begin{equation}
\label{function f fluid}
%\label{change moment of inertia total fluid}
%\frac{\Delta I_{ac}}{I}=\frac{\Delta\Omega^{2}_{ac}}{3\Omega^{2}_{K}}\,%\left(
\Delta\tilde I(\Delta\tilde\rho,\tilde t,0)=\frac{1+\Delta\tilde\rho(1+\tilde r_{i}^{5})+\Delta\tilde{\rho}^2\tilde r_{i}^{8}}{1+\Delta\tilde\rho(1+\tilde r_{i}^{5})+\Delta\tilde{\rho}^2\tilde r_{i}^{5}}.%\right).
\end{equation}
%Thus, evaluating 
Replacing in Eq.~(\ref{rigidity general}), one obtains a general expression for the rigidity parameter in a star with a uniform crust and a uniform core with possibly different densities,
\begin{eqnarray}
\label{rigidity evaluated}
 \lefteqn{
b(\Delta\tilde\rho,\tilde t,\beta)=} \nonumber\\
&&\beta[11\Delta\tilde\rho(d_{\Delta2}w_0^2+\tilde r_i^5 d_1-2\tilde r_i^5n_0w_0)+24 d_2 \tilde t \beta(w_0+\tilde r_i^5\Delta\tilde\rho)] \nonumber \\ &&[3d_{\Delta1}w_{0}\Delta\tilde{\rho}\tilde{t}+11d_{\Delta2}w_{0}\Delta\tilde{\rho}\beta+11d_{1}\beta+24d_{2}\tilde{t}\beta^{2}]^{-1} \nonumber \\
&& (w_0+\tilde r_i^5\Delta\tilde\rho)^{-1}.
%&&\equiv \frac{\beta \tilde t}{2(1+\Delta\tilde\rho)} \tilde b(\tilde r_i,\Delta\tilde\rho,\beta).
\end{eqnarray}
(A more explicit expression for $b$ is given in Eq.~(\ref{rigidity evaluated 2}).)
\begin{figure}
    \centering
    \resizebox{\hsize}{!}{\includegraphics{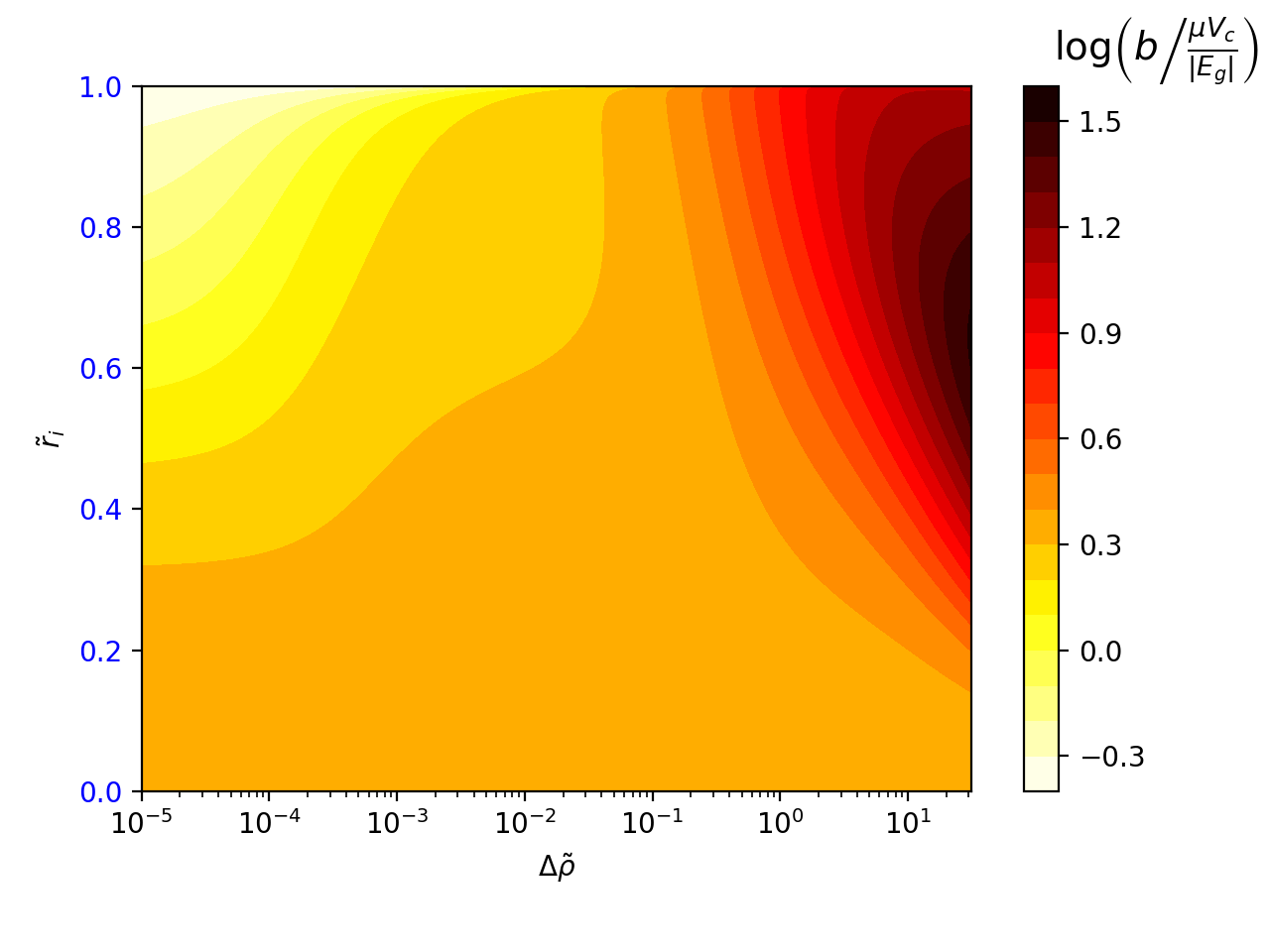}}
    \caption{Rigidity parameter $b/(\mu V_c/|E_{g}|)$ as a function of the difference in density between fluid core and solid crust $\Delta\tilde\rho$ in the horizontal axis and as a function of the core radius $\tilde r_i$ in the vertical axis. The fiducial parameters are fixed so that $\tilde \beta=\frac{4\pi\mu R^4}{3 G M^2}=0.8\times 10^{-5}$. }
    \label{fig:rigidity parameter}.
\end{figure}
%If we write the parameter $\beta \propto \mu R_{\star}^4 M_{\star}^{-2}$  or $\propto \mu V_c |E_g^{-1}|$ , where 
%, this $\beta$ parameter will be a different function of $\tilde r_i$ and $\Delta \tilde \rho$ as

It can be useful to rewrite the dimensionless elasticity parameter $\beta$ in two different ways:
\begin{eqnarray}
\label{beta parameter}
    \beta& =%\equiv
    & \tilde\beta(1+\Delta\tilde\rho\tilde r_i^3) 
    %=\frac{8\pi\mu R_{\star}^4}{3 G M_{\star}^2}(1+\Delta\tilde\rho\tilde r_i^3)  
    \\
    %&=&1.6\times10^{-5}(1+\Delta\tilde\rho\tilde r_i^3)\frac{\mu}{10^{30}\frac{\mathrm{erg}}{\mathrm{cm}^{3}}}\left(\frac{R_{\star}}{10^{6}\mathrm{cm}}\right)^4\left(\frac{1.4 M_{\odot}}{M_{\star}}\right)^2 \nonumber\\
    &=& \frac{\mu V_c}{|E_g|}\frac{[1+\tilde r_i^3\Delta\tilde\rho(\tilde r_i^2+3)/2+\tilde r_i^5\Delta\tilde\rho^2]}{5\gamma_2\tilde t(1+\tilde r_i^3\Delta\tilde\rho)}, 
    %&=& 1.1 \times 10^{-5} \tilde \beta(\Delta\tilde\rho,\tilde r_i) \frac{\mu}{10^{30}\frac{\mathrm{erg}}{\mathrm{cm}^{3}}}\left(\frac{V_c}{1.3 \times 10^{17}\mathrm{cm}^3}\right)\left(\frac{1.1 \times 10^{53}\mathrm{erg}}{|E_{g}|}\right)^2 \nonumber.
\end{eqnarray}
where 
\begin{equation}
\label{beta tilde}
    \tilde \beta%=
    \equiv\frac{4\pi\mu R_{\star}^4}{3 G M_{\star}^2}=0.8\times10^{-5}\frac{\mu}{10^{30}\frac{\mathrm{erg}}{\mathrm{cm}^{3}}}\left(\frac{R_{\star}}{10^{6}\mathrm{cm}}\right)^4\left(\frac{1.4 M_{\odot}}{M_{\star}}\right)^2,
\end{equation}
$V_c=4\pi R_{\star}^3(1-\tilde r_i^3)/3=4\pi R_{\star}^3\gamma_2\tilde t$ is the volume of the crust, 
\begin{equation}
    \label{gravitational binding energy}
    E_g=E_{g,0}\frac{1+\Delta\tilde\rho\tilde r_i^3(\frac{\tilde r_i^2}{2}+\frac{3}{2})+\Delta\tilde\rho^2\tilde r_i^5}{(1+\Delta\tilde\rho\tilde r_i^3)^2},
\end{equation}
is the gravitational binding energy of our model star, and $E_{g,0}=-3GM_{\star}^2/(5R_{\star})$ is the gravitational binding energy of a uniform sphere of the same mass and radius.

Figure~\ref{fig:rigidity parameter} shows that for all values of $\tilde r_i$ and for a density contrast in the range $0\leq\Delta\tilde\rho\lesssim 30$, the rigidity parameter $b$ does not differ from the dimensionless combination $\mu V_c/|E_g|$ by more than an order of magnitude. %\footnote{However, for unrealistic combinations of parameters, $b$ can become much larger, e.~g., for $\tilde\beta\sim 10^{-5}$, $\Delta\tilde\rho\sim 10^7$, and $\tilde r_i\sim 0.1$, we obtain $b\sim 0.3$ and $b/(\mu V_c/|E_g|)\sim 10^5$.}. 
%The full analytical expression for the ratio $b/(\mu V_c/|E_g|)$ in terms of $\Delta\tilde\rho$, $\tilde\beta$, and $\tilde r_i$ is given in equation~(\ref{rigidity evaluated 2}).

The horizontal axis of Fig.~\ref{fig:rigidity parameter}, %origin, 
$\tilde r_i=0$, %and $\Delta\tilde\rho=0$, is the solid uniform case of
corresponds to the case of a uniform elastic sphere studied by \citet{1971AnPhy..66..816B}.
%, which remains constant along all the coordinate axes. 
In this case, we obtain $\beta=\tilde\beta$ and
\begin{equation}\label{rigidity uniform elastic}
b=\frac{\frac{19}{5}\tilde\beta}{1+\frac{19}{5}\tilde\beta}\approx
%\frac{76\pi\mu \,R_{\star}^4}{15 G \,M_{\star}^2}=
\frac{57}{25}\frac{\mu V_T}{|E_{g0}|}, %\\
\end{equation}
where $V_T$ is the total volume of the star, and the %latter 
approximation corresponds to the realistic limit $\tilde\beta=(3/5)\mu V_T/|E_g|\ll 1$. 
The vertical axis, corresponding to $\Delta\tilde\rho\to 0$, approaches the case studied by \citet{2000ApJ...543..987F}, namely a star with a uniform fluid core and a uniform elastic crust, both with the same density, for which  Eq.~(\ref{rigidity evaluated}) with $\beta=\tilde\beta\ll 1$ yields
\begin{equation}
\label{rigidity parameter uniform star fluid core}
b\approx\frac{24 d_2}{55\gamma_2 d_1}\frac{\mu V_c}{|E_g|}\approx\frac{24}{55}\frac{\mu V_c}{|E_{g}|},
\end{equation} 

where the latter approximation also assumes $\tilde t\ll 1$, making $d_1=d_2=\gamma_2=1$. 

A somewhat more general expression can be obtained by taking simultaneously $\Delta\tilde\rho\ll 1$, $\tilde\beta\ll 1$, and $\tilde t\ll 1$ in Eq.~(\ref{rigidity evaluated 2}). This allows us to set $w_0=1$ and $\beta=\tilde\beta$, keep only the lowest-order terms in $\Delta\tilde\rho$ and $\tilde\beta$ in the numerator and the denominator, and keep the lowest-order term in $\tilde t$ in each of their coefficients, which yields
\begin{equation}\label{rigidity triple approx}
 b=3\tilde\beta\tilde t\,\frac{9\Delta\tilde\rho\tilde t+8\tilde\beta}{3\Delta\tilde\rho\tilde t+11\tilde\beta}.   
\end{equation}
Thus, as long as $\Delta\tilde\rho\tilde t\ll\tilde\beta\ (\ll 1)$, we recover the result of Eq.~(\ref{rigidity parameter uniform star fluid core}), whereas for $\tilde\beta\ll\Delta\tilde\rho\tilde t\ (\ll 1)$ we have 
\begin{equation}\label{rigidity beta=0}
    b\approx 9\tilde\beta\tilde t%-25\frac{\beta^2}{\Delta\tilde\rho} 
    \approx\frac{9}{5}\frac{\mu V_c}{|E_g|}.
\end{equation}
The change of behavior near $\Delta\tilde\rho\sim\tilde\beta/\tilde t$ can be clearly seen in Fig.~\ref{fig:rigidity parameter}. To the right of this point, the density contrast is still small, but the elasticity parameter $\tilde\beta$ is even smaller, corresponding to taking first the limit $\tilde\beta\to 0$ and then $\Delta\tilde\rho\to 0$, that is, in the opposite order than in Eq.~(\ref{rigidity parameter uniform star fluid core}). 
%which shows than $\Delta\tilde\rho$ and $\beta$ does not commute when they both goes to zero. This could be because i
If the limit $\tilde\beta\to 0$ is taken first, the crust-core interface can be viewed as separating two fluids of different densities, so the interface is roughly an equipotential. %In the opposite case
If, instead, the limit $\Delta\tilde\rho\to 0$ is taken first, the crust-core interface separates a fluid and an elastic solid of the same density, so the magnitude of its %shape %is determined by 
deformation depends crucially on the shear modulus of the solid. 

Since, in a real neutron star, $\tilde\beta$ is very small ($\sim 10^{-5}$, see Eq.~(\ref{beta tilde})), the crust is relatively thin ($\tilde t\sim 0.1\ll 1$), but the contrast between the average crust and core densities is in the range $10\lesssim\Delta\tilde\rho\lesssim 30$, we consider the limit of $\tilde\beta\to 0$ and $\tilde t\to 0$ but finite $\Delta\tilde\rho$ (roughly in the range $\tilde\beta/\tilde t\sim 10^{-4}\ll\Delta\tilde\rho\ll\tilde t/\tilde\beta\sim 10^4$), for which we obtain
\begin{equation}
\label{rigidity big dro}
\frac{b}{\mu V_c/|E_g|} =\frac{3(3+16\Delta\tilde\rho+24\Delta\tilde\rho^2)}{5(1+\Delta\tilde\rho)^2}.
\end{equation}
As the density contrast increases, this function increases from $9/5=1.8$ at  $\Delta\tilde\rho=0$ toward an asymptotic limit of $72/5=14.4$ for $\Delta\tilde\rho\to\infty$. We note that these values exceed the one obtained for $\Delta\tilde\rho=0$ and $0<\tilde\beta\ll 1$ (Eq.~(\ref{rigidity parameter uniform star fluid core})) by a factor between $33/8$ and $33$. For $10<\Delta\tilde\rho<30$, we obtain $12.7<b/(\mu V_c/|E_g|)<13.8$, not far from the asymptotic limit. Taking  $b/(\mu V_c/|E_g|)\approx 13$, noting that in this regime $\mu V_c/|E_g|\approx 5\tilde\beta\tilde t$, with $\tilde\beta$ given by Eq.~(\ref{beta tilde}), and applying a correction factor of $5/2$ to account for the effect of gravitational potential perturbations (see Sec.~\ref{sec:Cowling}), we obtain our best estimate for the rigidity parameter of a neutron star, based on our model with a uniform crust and uniform core of different densities,
\begin{equation}\label{best estimate b}
    b\approx 1.3\times 10^{-4}\frac{\mu}{10^{30}\mathrm{erg\,cm}^{-3}}\frac{\Delta R}{1\mathrm{km}}\left(\frac{R_{\star}}{10\mathrm{km}}\right)^3\left(\frac{1.4 M_{\odot}}{M_{\star}}\right)^2.
\end{equation}

\subsection{Effect of the Cowling approximation} \label{sec:Cowling}

Both our results for an uniform, elastic star (Eq.~(\ref{rigidity uniform elastic})) and for a star with an elastic crust and a fluid core of equal, uniform densities (Eq.~(\ref{rigidity parameter uniform star fluid core})) are lower by a factor of $2/5$ than their counterparts found in the literature, namely the classical result of Lord Kelvin (reproduced by \citealt{1971AnPhy..66..816B} and \citealt{2003ApJ...588..975C}) for the first case, and the result of \citet{2003ApJ...588..975C} for the model of \citet{2000ApJ...543..987F} for the second. 

This factor can be understood from the calculations by \citet{2019PASA...36...36G} of the displacement fields for a star with an elastic crust and a solid core. From their Eqs. (10) and (57), the quadrupolar radial displacement field can be written as
\begin{equation}\label{displacement delta}
    \xi^r_{2,\delta}=\frac{P(\tilde t, r)}{Q(\tilde t)+60\tilde\beta\tilde t}
\end{equation}
when gravitational potential perturbations are taken into account, and
\begin{equation}\label{displacement Cowling}
    \xi^r_{2,C}=\frac{2}{5}\frac{P(\tilde t, r)}{Q(\tilde t)+24\tilde\beta\tilde t}
\end{equation}
when they are not, that is, in the Cowling approximation. In both equations, $P$ and $Q$ are functions whose precise form is unimportant for the present argument, and we have neglected higher-order terms in $\tilde t$ and $\tilde\beta$ ($q$ and $\chi^2$, in the notation of \citealt{2019PASA...36...36G}), both of which are assumed to be small. From our Eqs.~(\ref{change moment of inertia total}) and (\ref{rigidity general}), we see that the rigidity parameter depends on the fractional difference between the displacements for a purely fluid star ($\tilde\beta=0$) and a star with an elastic crust ($\tilde\beta\neq 0$),
\begin{equation}
    b\propto 1-\frac{\xi^r_2(\tilde\beta\neq 0)}{\xi^r_2(\tilde\beta=0)}.
\end{equation}
For the case with gravitational potential perturbations, the right-hand side of this equation is $\approx 60\tilde\beta\tilde t/Q(\tilde t)$, whereas in the Cowling approximation it is $\approx 24\tilde\beta\tilde t/Q(\tilde t)$; therefore, the ratio of the rigidity parameters is, to lowest order in $\tilde\beta$ and $\tilde t$,
\begin{equation}
    \frac{b_C}{b_\delta}=\frac{24}{60}=\frac{2}{5},
\end{equation}
confirming the factor found in our comparison. 

We note that this ratio $2/5$ does not come from the overall prefactor in Eq.~(\ref{displacement Cowling}) but from the ratio of the coefficients (24 and 60) of the small correction terms of order $\tilde\beta\tilde t$. It is by no means clear whether in the more general case with different densities this ratio is still %has a constant factor of 
$2/5$ or whether it %this factor 
now depends on the density contrast $\Delta\tilde\rho$. We can only guess that the ratio might not be very different, so our results for the rigidity parameter $b$ in Sec.~\ref{sec:rigidity} should all be multiplied by a factor of $\approx 5/2$ in order to account for gravitational potential perturbations (except the final Eq.~(\ref{best estimate b}), where it is already taken into account).

\subsection{Recovery fraction and glitch activity} \label{sec:recovery}

\begin{figure}
\centering
\resizebox{\hsize}{!}{\includegraphics{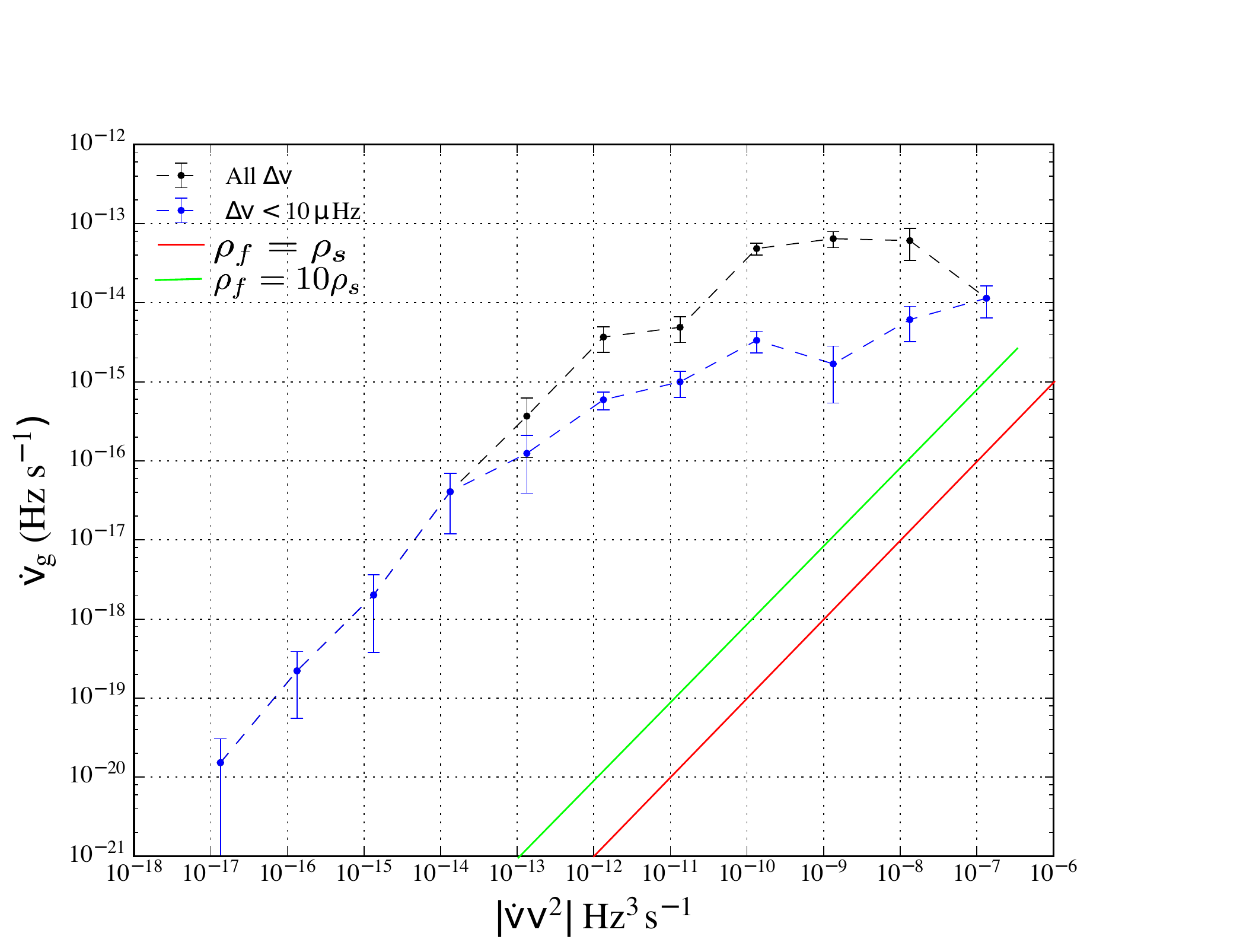}}\\
\caption{Glitch activity $\dot{\nu}_{g}$ as a function of $\dot{\nu}\nu^{2}$. Black dots represent the glitch activity (sum of all glitch sizes divided by the sum of the observation times) for all pulsars within bins of $\log(\dot{\nu}\nu^{2})$, and blue dots represent the same quantity but restricted to small glitches ($\Delta\nu<10\mathrm{\mu Hz}$). (Data kindly provided by J. R. Fuentes, from the glitch sample discussed in \citealt{2017A&A...608A.131F}.) The green and red lines represent %upper bounds regarding the maximum "$b$" that can be chosen for 
the glitch activity in the starquake model, as given by Eq.~(\ref{glitch activity}), for the values $b=10^{-5}$ and $b=10^{-6}$ of the rigidity parameter, corresponding to the cases $\Delta\tilde\rho=9$ and $\Delta\tilde\rho=0$, respectively. % $\dot\nu_g/(\nu^2|\dot\nu|)=2\times10^{-13}(b/4\times10^{-6})$. 
}.
\label{fig:rafaelplot2.pdf}
\end{figure}

Combining Eqs.~(\ref{glitch conservation}), (\ref{change moment of inertia total}), and (\ref{rigidity general}), we obtain the ratio of the spin-up in a glitch to the spin-down between glitches,
\begin{equation}
\label{recover from glitch 2}
\frac{\Delta\Omega_{bc}}{\Delta\Omega_{ac}}=\frac{2}{3}b\left(\frac{\Omega }{\Omega_{K}}\right)^{2}\Delta\tilde I(\Delta\tilde\rho,\tilde t,0),
\end{equation}
valid for all the cases, where $\Delta\tilde I(\Delta\tilde\rho,\tilde t,0)$, given by Eq.~(\ref{function f fluid}), decreases from %the normalized variation of the moment of inertia of the fluid lies between 
$1$ for $\Delta\tilde\rho=0$ to $\tilde r_i^3\sim 0.7$ for $\Delta\tilde\rho\to\infty$, with $\Delta\tilde I(\Delta\tilde\rho,\tilde t,0)\approx 0.8$ in the range $10<\Delta\tilde\rho<30$, which we adopt as our standard value in what follows. The ratio given in Eq.~(\ref{recover from glitch 2}) is particularly interesting, because both the numerator and the denominator on the left-hand side are measurable in observed glitches. If all glitches were equal,

we could identify
\begin{equation}
\label{ratio spin down-glitch}
\frac{\Delta\Omega_{bc}}{\Delta\Omega_{ac}}=\frac{\dot{\nu}_{g}}{|\dot{\nu}|},
\end{equation} 
where $\nu=\Omega/(2\pi)$ is the rotation frequency, $\dot{\nu}$ is its time-derivative (the spin-down rate of the pulsar), 
\begin{equation}
\dot{\nu}_{g}\equiv\frac{\sum_{i}\Delta\nu_{i}}{T}
\end{equation} 
is the glitch spin-up rate, customarily called ``glitch activity'' \citep{2000MNRAS.315..534L}, which accounts for the cumulative spin-up effect of glitches during the time of observation $T$ of a given pulsar, and $\Delta\nu_i$ is the size of its $i$-th glitch observed during this time. Therefore, from Eq.~(\ref{recover from glitch 2}), the glitch activity in the starquake model can be written as
\begin{eqnarray}
\label{glitch activity}
\frac{\dot{\nu}_{g}}{|\dot{\nu}|} &\approx& 0.53\left(\frac{\nu}{\nu_K}\right)^2 b \\
&\approx& %2.4\times 10^{-11} 
1.4\times 10^{-5}\left(\frac{\nu}{\mathrm{kHz}}\right)^{2}\frac{\mu}{10^{30}\frac{\mathrm{erg}}{\mathrm{cm}^{3}}}\frac{\Delta R}{1\mathrm{km}}\left(\frac{R_{\star}}{10\mathrm{km}}\right)^6\left(\frac{1.4 M_{\odot}}{M_{\star}}\right)^3, \nonumber%\mathrm{s^{2}},
\end{eqnarray}
where $\nu_{K}=(GM_{\star}/R_{\star}^{3})^{1/2}/(2\pi)\approx 2.2\,\mathrm{kHz}$ is the Keplerian frequency. We note that all glitching pulsars have spin frequencies substantially lower than 1 kHz (e.g., $\nu\approx 11\,\mathrm{Hz}$ for Vela), whereas many (perhaps most) have measured glitch activities $\dot\nu_g/|\dot\nu|\approx 0.01$ \citep{2000MNRAS.315..534L,2011MNRAS.414.1679E,2017A&A...608A.131F}. Thus, starquakes (crust breaking events) are clearly far from being able to account for all glitches. Figure \ref{fig:rafaelplot2.pdf} shows that the observed glitch spin-up rates in most bins are 5-6 orders of magnitude higher than the prediction based on starquakes, clearly indicating that this model cannot account even for the subsample of small glitches ($\Delta\nu<10\,\mathrm{\mu Hz}$) identified from the observed bimodality in glitch sizes \citep{2011MNRAS.414.1679E,2017A&A...608A.131F}. We emphasize that this result is independent of the rather uncertain breaking strain of the neutron star crust material, which determines the size and frequency of starquakes but without affecting $\dot\nu_g$.

\subsection{Crust breaking, glitch size, and glitch frequency}

The two most widely used criteria for the failure of solid materials are 
the von Mises criterion, 
\begin{equation}
\label{von Mises criterion}
\sigma_M^2\equiv\frac{1}{6}\left[(\sigma_1-\sigma_2)^2+(\sigma_1-\sigma_3)^2+(\sigma_2-\sigma_3)^2\right]\geq \sigma_{M,c}^2,
\end{equation} 
and the Tresca criterion,
\begin{equation}
\label{Tresca criterion}
\sigma_T\equiv\frac{1}{2}|\sigma_1-\sigma_3|\geq \sigma_{T,c},
\end{equation}
where $\sigma_1\geq\sigma_2\geq\sigma_3$ are the eigenvalues of the strain tensor $\tens{\sigma}$,%\footnote{Alternatively, $\alpha\equiv|\sigma_1-\sigma_3|=2\sigma_T$ %(without the prefactor $1/2$) 
%is also defined as the \emph{strain angle} (e.~g., \citealt{2000ApJ...543..987F}).} %(equivalent to the shear strain tensor $\tens\Sigma$ in our incompressible model, in which $\sigma_1+\sigma_2+\sigma_3=0$), 
and $\sigma_{M,c}$, $\sigma_{T,c}$ are dimensionless ``critical strains'', a property of the material that is difficult to predict from theory \citep{christensen2013theory}. These two criteria can be compared by considering $\sigma_M^2$ as a function of the intermediate eigenvalue $\sigma_2$. This function is represented by a parabola with minimum at $\sigma_2=(\sigma_1+\sigma_3)/2$, at which $\sigma_M=\sigma_T$. On the other hand, at the extremes of the interval $\sigma_3\leq\sigma_2\leq\sigma_1$, one obtains $\sigma_M=2\sigma_T/\sqrt{3}$, and  therefore
\begin{equation}    1\leq\frac{\sigma_M}{\sigma_T}\leq 1.155,
\end{equation}
making the two criteria essentially equivalent if $\sigma_{M,c}\approx\sigma_{T,c}$, so we chose to work with the mathematically more tractable von Mises criterion. 

\begin{figure}
\resizebox{\hsize}{!}{\includegraphics[width=.5\textwidth,center]{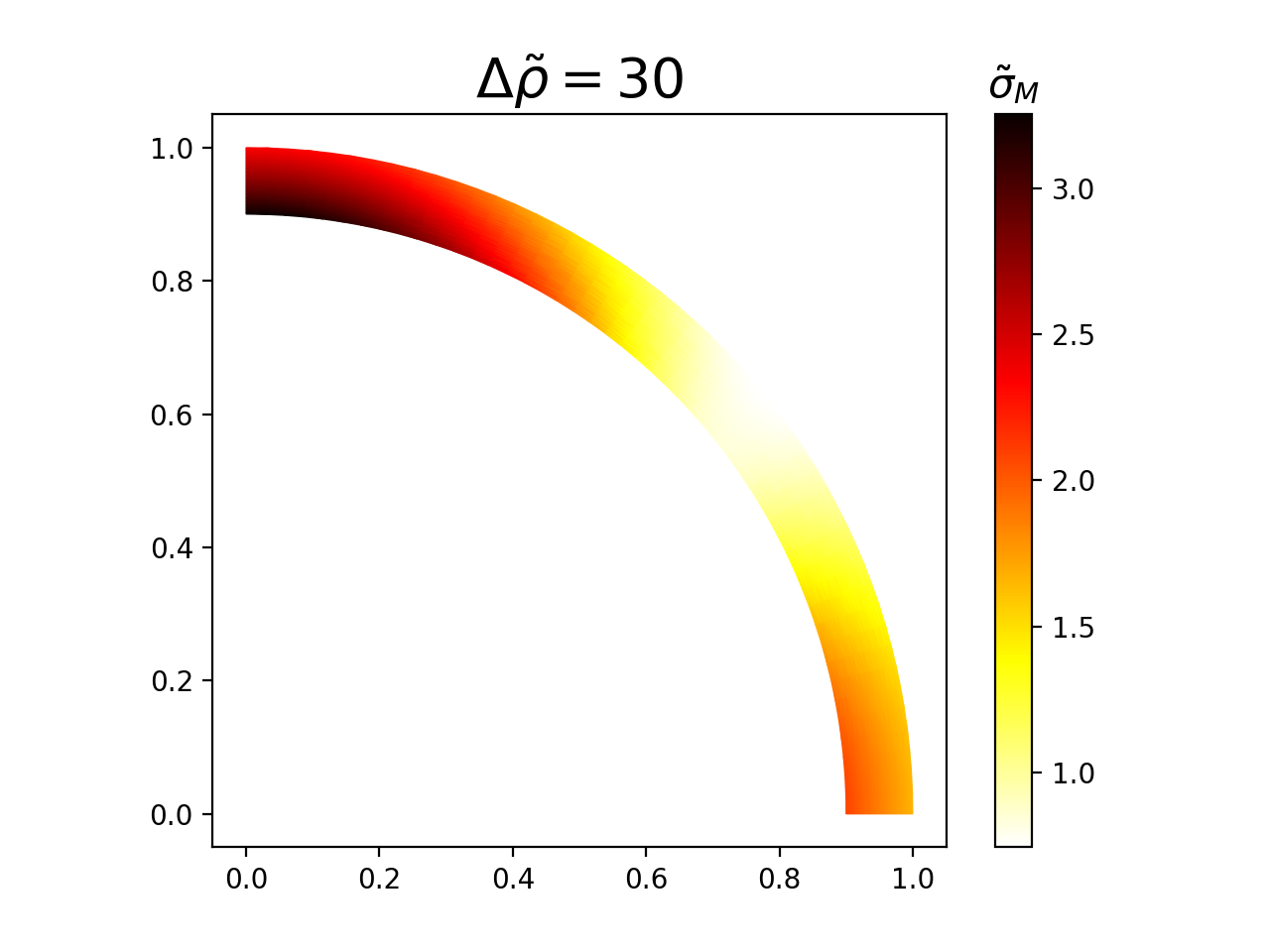}}
\caption{Normalized von Mises strain, $\tilde\sigma(\tilde r,\theta)$ (see Eqs.~(\ref{von Mises criterion}) and (\ref{von Mises 3})), for $\tilde r_{i}=0.9$, $\Delta\tilde{\rho}=30$, and $\tilde\beta=0.8\times10^{-5}$.}
\label{fig:von Mises criterion}
\end{figure}

In incompressible matter, the strain tensor is traceless; therefore $\sigma^{ij}=\Sigma^{ij}$, and we can also write
\begin{equation}
\label{von Mises criterion2}
\sigma_M^2=\frac{1}{2}(\sigma_1^2+\sigma_2^2+\sigma_3^2)=\frac{1}{2}\Sigma^{ij}\Sigma_{ij}.
%=\frac{\delta\phi_{2}}{R_{\star}g(R_{\star})}\sqrt{\frac{1}{2}\tilde{\Sigma}^{ij}\tilde{\Sigma}_{ij}}\equiv \frac{\delta\Omega^{2}}{3\Omega_{K}^{2}}\tilde{\sigma}.
\end{equation}
In our %neutron star 
model, this expression can be evaluated using the equations in Sects. \ref{sec:elastic} and \ref{sec:two domains}, and rewritten as
\begin{equation}
\label{von Mises 3}
\sigma_M=\frac{\Delta(\Omega^{2})}{3\Omega_{K}^{2}}\tilde\sigma_M(r,\theta),
\end{equation}
where $\tilde\sigma_M(r,\theta)$ is a dimensionless function of position in the crust, shown in Fig.~\ref{fig:von Mises criterion}. %and analyzed in detail in appendix \ref{section:Strains due to rotation}.  

\begin{figure*}
\resizebox{\hsize}{!}{\includegraphics{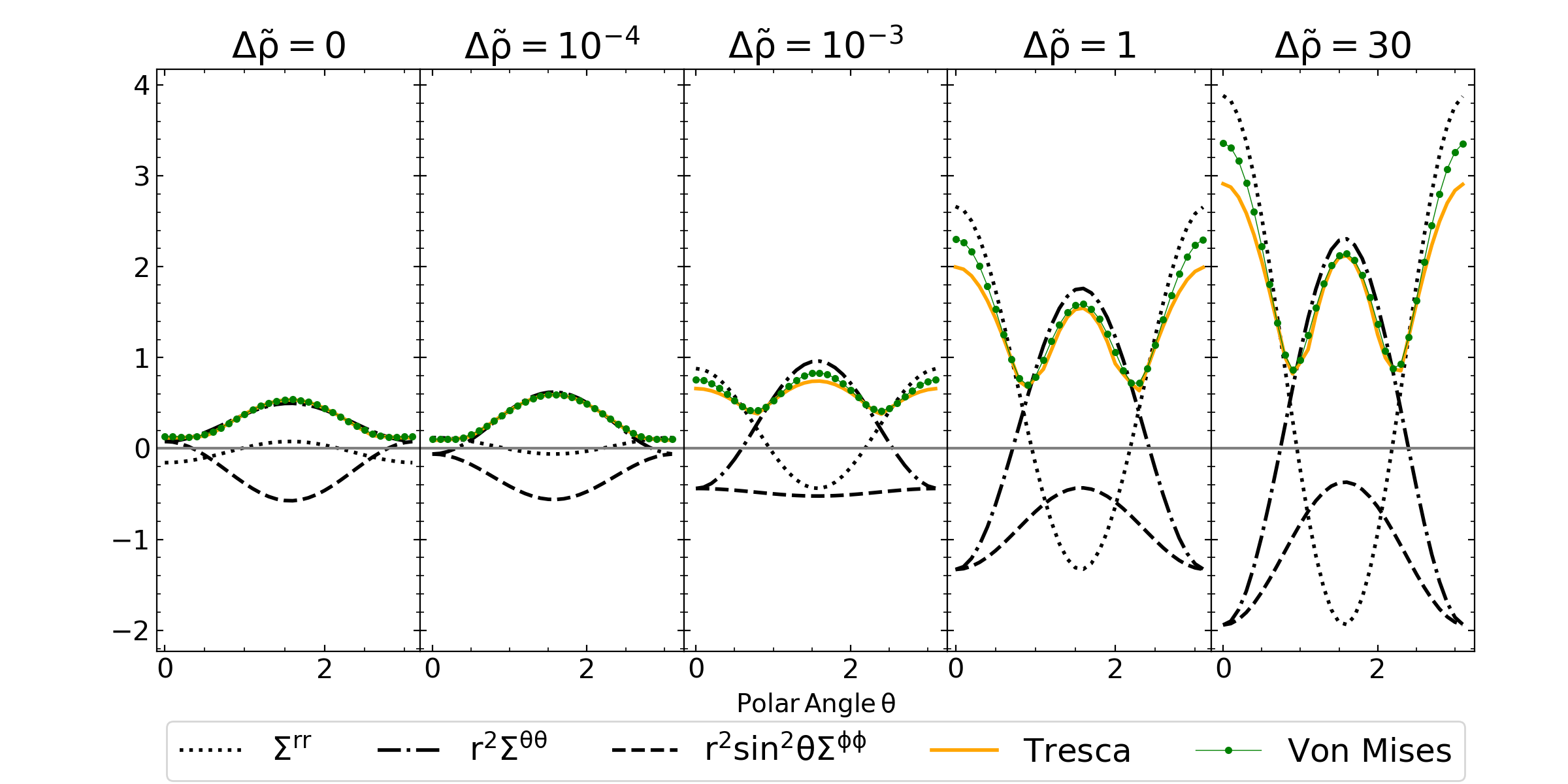}}
    \caption{Strain components, von Mises strain $\sigma_M$ (Eq. (\ref{von Mises criterion})), and Tresca strain $\sigma_T$ (Eq. (\ref{Tresca criterion})) %Tresca criterion (which is half of the strain angle $\sigma_{T}=\alpha/2=(\sigma_{max}-\sigma_{min})/2$), and "norm" of the strain tensor from von Mises criterion  
    at the crust-core interface for %of the star with 
    $\tilde r_{i}=0.9$ and $\tilde\beta=10^{-5}$. %, produced by the centrifugal force. 
    From left to right, %: first panel of core and crust with same density, to the right 
    the dimensionless density contrast increases as labeled. %up to $\Delta\tilde\rho=10$ in the last panel.  %The panel of the right shows the strains for a star with a core equally denser its crust, meanwhile in the left panel the core is 11 times denser than its crust
    }
    \label{fig: Tresca and von Mises}
\end{figure*}

By analogy with solids on Earth (at essentially zero pressure) and attempting to match the observed glitch phenomenology, it was originally assumed that the critical strains of neutron star crusts were very small, $\lesssim 10^{-4}$ %or even smaller 
\citep{1970PhRvL..24..923S}. However, %recent 
molecular dynamics simulations \citep{2009PhRvL.102s1102H,2012MNRAS.426.2404H} of solids subject to very high pressures, as in neutron star crusts, support much larger values. \citet{2009PhRvL.102s1102H} use a Cartesian box, applying an incompressible deformation\footnote{For $\xi^z$ there is a misprint in \citet{2009PhRvL.102s1102H}, but this does not affect our results. The correct expressions are given in \citet{2008arXiv0812.2650H}.} $\xi^x=\alpha y/2$, $\xi^y=\alpha x/2$, $\xi^z={\cal O}(\alpha^2)$, and finding that the material yields when $\alpha\approx 0.1$, corresponding to $\sigma_{M,c}=\sigma_{T,c}=\alpha/2\approx 0.05$ according to our definitions above\footnote{$\alpha$ is the "strain angle", also used by \citet{2000ApJ...543..987F}.}. %On the other hand, 
\citet{2018MNRAS.480.5511B} find that the breaking strain can vary by a factor of $\sim 3$ with respect to the result of \citet{2009PhRvL.102s1102H}, depending on the direction of the deformation. However, given the many other uncertainties involved in this problem, we use the constant value of \citet{2009PhRvL.102s1102H} as our reference. % argue than the critical strain for contraction of the material can be 2.5 smaller than the breaking strain of \citet{2009PhRvL.102s1102H}, reducing the critical strain to $\sigma_{crit}\approx 2\times 10^{-2}$.

%In the incompressible star, %in addittion to the smallness of 
Since the off-diagonal strain components $\Sigma^{r\theta}$ are forced to vanish at both the stellar surface ($\tilde r=1$) and the crust-core interface ($\tilde r=1-\tilde t$), they are generally much smaller (by a factor of $\sim\tilde t^2\ll 1$) than    %where 
%\begin{eqnarray}
%\xi^{\theta}=-\left(\frac{\xi^r_2}{2r}+\frac{\partial_r\xi^r_2}{4}\right)\sin 2\theta,
%\end{eqnarray}
the diagonal components, which can be written as
\begin{eqnarray}
\Sigma^{rr}&=&\partial_r\xi^r_2 \left(1-\frac{3}{2}\sin^2\theta\right), \\
r^2\Sigma^{\theta\theta}&=&-\frac{\partial_r\xi^r_2}{2}+\left(\partial_r\xi^{r}_2+\frac{\xi^r_2}{2r}\right)\sin^2\theta,  \\
r^2\sin^2\theta\Sigma^{\phi\phi}&=&-\frac{\partial_r\xi^r_2}{2}+\frac{1}{2}\left(\partial_r\xi^{r}_2-\frac{\xi^r_2}{r}\right)\sin^2\theta.
\end{eqnarray}
%they are very close to the principal components if the thickness of the crust is small with a difference of $\tilde \Sigma^{r\theta}\ll \tilde t$ regarding the principal components. So, with some algebra, 
Thus, ignoring the off-diagonal terms, the squared von Mises strain becomes
\begin{eqnarray}
\sigma_{M}^2(r,\theta)&\approx&\frac{3}{2}(\partial_r\xi^r_2)^2\left(1-3\sin^2\theta\right) %\times 
\nonumber \\ %\times 
& &+\frac{1}{2}\left[7(\partial_r\xi^r_2)^2+\frac{\xi^{r}_2}{r}\partial_r\xi^r_2+ \left(\frac{\xi^{r}_2}{r}\right)^2 \right] \sin^4\theta,
\end{eqnarray}
which is a decreasing function of $\sin\theta$ for small values of $\sin\theta$, and an increasing function for $\sin\theta\sim 1$. Therefore, its maximum value, where the crust first breaks, must lie either at the equator ($\theta=\pi/2$), which happens for $-\xi^r_2/(2r)<\partial\xi^r_2/\partial r<\xi^r_2/r$, or at the pole ($\theta=0$), which happens outside this range.

%In the uniform situation 
For very small $\Delta\tilde\rho$ (the condition, for $\beta\ll\tilde t\ll 1$, is $\Delta\tilde\rho<4\beta/[3\tilde t]\sim 10^{-4}$), the radial displacement function $\xi^r_2(r)$ has a maximum inside the star and decreases slightly from the interface to the surface (upper panel of Fig.~\ref{fig:crust}), so %red line of top panel of Fig. \ref{fig:rdispsurfint2}), so %the radial displacement at the interface, 
$\xi^r_2(\tilde r_i)>%$, is \emph{larger} than that at the surface %, 
\xi^r_2(R_{\star})$, %=\Delta(\Omega^2)/3\Omega^2_K$ (by a factor $1+4\tilde t/11$ in the limit $\Delta\tilde\rho=0$; see eq. \ref{radial displacement interface app FLE}), which implies that 
implying $\Sigma^{rr}_2=\partial\xi^r_2/\partial r < 0$. Thus, as the star reduces its equatorial bulge under spin-down,  the crust becomes thicker %radially expands 
at the equator and thinner %is compressed 
at the poles. As shown in the first panel of Fig.~\ref{fig: Tresca and von Mises}, the strain is mostly due to the nonradial components and is maximum at the equator, where the crust is expected to break first. %Also, the polar displacement $\xi^{\theta}=-(\xi^r_2/2)\sin2\theta$ is minimum in the uniform star because of the small radial shear strain. 

For $\Delta\tilde\rho\approx 4\beta/(3\tilde t)\sim 10^{-4}$, %10^{-3}$ and $\tilde r_i=0.9$, 
the radial displacement is %almost %constant
nearly independent of $r$, so $\Sigma^{rr}\approx 0$, and the crust neither expands nor shrinks radially at any latitude. The total strain is due to the polar and azimuthal components,
%\begin{eqnarray}
%\label{polar azimuthal strain uniform}
$r^2\Sigma^{\theta\theta}%(\Delta\tilde\rho=10^{-3})=
\approx -r^2\sin^2\theta\Sigma^{\phi\phi}\approx\xi^{\theta}_2(r)\sin^2\theta$, 
%\end{eqnarray}
%produce the total strain in the uniform star, which 
whose $\sin^2\theta$ dependence implies that the crust will first break at the equator (see Fig.~\ref{fig: Tresca and von Mises}, second panel).

\begin{figure}
    \centering
     \resizebox{\hsize}{!}{\includegraphics{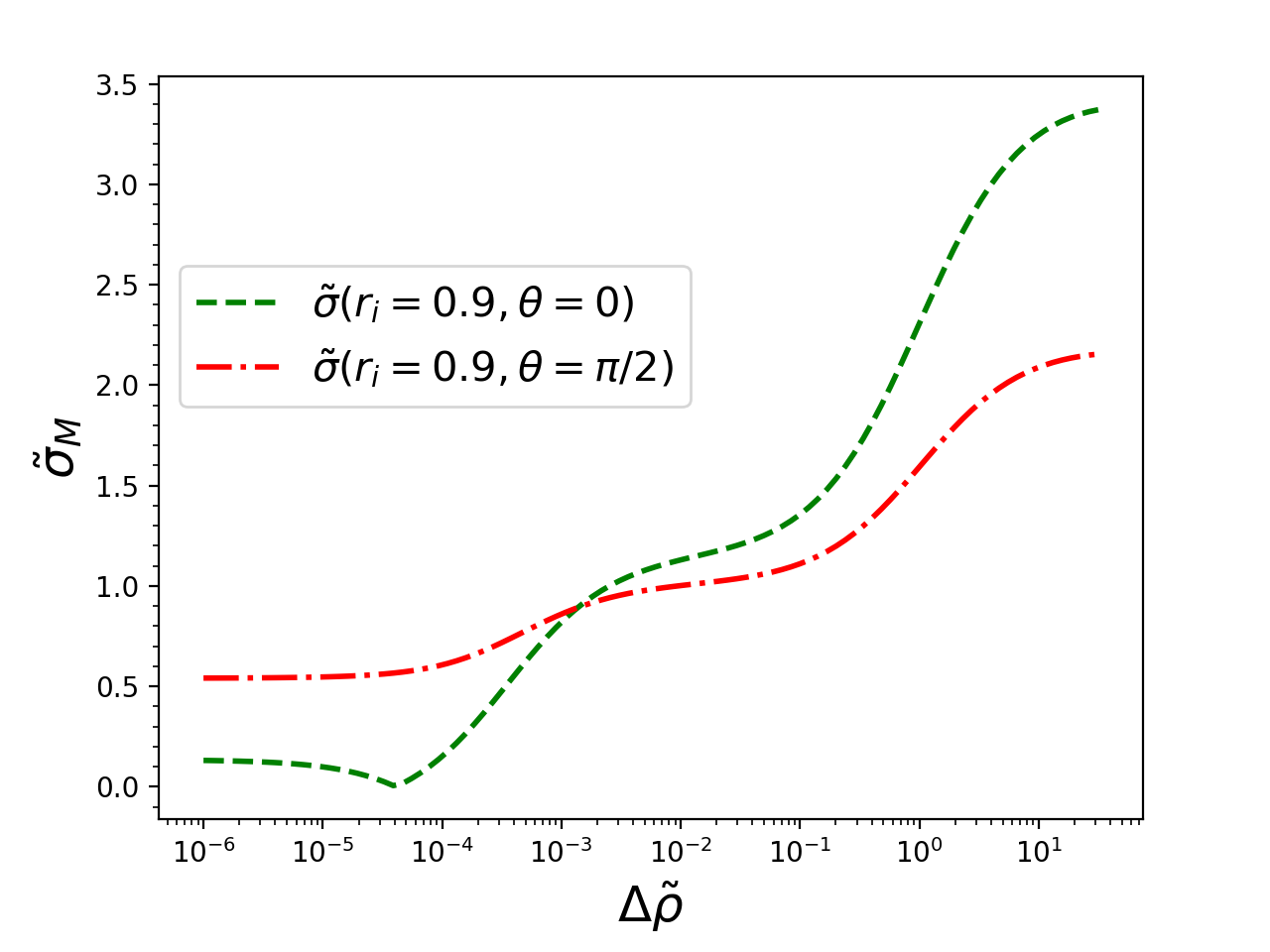}}
    \caption{Normalized von Mises strain $\tilde\sigma_M$ on the crust-core interface at the poles (green line) and at the equator (red line) as a function of the dimensionless density contrast $\Delta\tilde\rho$, with $\tilde\beta=10^{-5}$ and dimensionless crust thickness $\tilde t=0.1$. %, so that the von Mises strains are computed at the interface in this plot. %
    }
    \label{fig: strain maximum}
\end{figure}

The condition $\partial\xi^r_2/\partial r>\xi^r_2/r$, for which the largest strain occurs at the poles, is satisfied for $\Delta\tilde\rho\gtrsim[5\beta/(3\tilde t)]^{1/2}\sim 10^{-2}$, which includes the arguably most realistic case, $\Delta\tilde\rho\gtrsim 1$, in which the interface and the surface are roughly equipotentials, as for a completely fluid star. In this situation, the radial displacement increases almost linearly from the interface to the surface %(eqs. [\ref{radial displacement interface solid}], [\ref{radial displacement surface solid}], and 
(lower panel of Fig.~\ref{fig:crust}), so $\Sigma^{rr}_2\approx\mathrm{constant}>0$,
implying that the crust compresses radially at the equator and expands at the poles.
Figure~\ref{fig: strain maximum} shows that the maximum strain, hence the first breaking point, occurs at the poles. %for a neutron star with the core a few times denser than the crust.

%As we can see in 
Figures~\ref{fig:von Mises criterion} and %in Fig. [
\ref{fig: strain maximum} show that the maximum value of $\tilde{\sigma}$ for $\rho_f\approx 10 \rho_s$ is $\tilde\sigma_{max}\approx 3$, obtained at the poles of the crust-core interface, where the criterion indicates that fractures will begin. 
%, the total equatorial bulge decreases but the crust in there expands. Besides, s
Since $\beta$ is small ($\lesssim 10^{-4}$), its precise value has no relevant influence on the shear strain. %because it is in the denominator of \ref{constants displacement2} and \ref{constants displacement3}. 
%Thus, %the von Mises criterion (eq. \ref{von Mises criterion}) implies that 
From Eqs. (\ref{von Mises 3}), %we obtain that 
the change in the squared rotation rate required for an initially relaxed star to produce a starquake is 
\begin{equation}
\label{von Mises 4}
\frac{\Delta(\Omega^2)_{ab}}{\Omega_K^2}=0.1\left(\frac{\sigma_c}{0.1}\right)\left(\frac{3}{\tilde\sigma_{M,max}}\right),
\end{equation}
where $\sigma_c$ is a generic critical strain, for whatever criterion is applied. Thus, for the high critical strains obtained by \citet{2009PhRvL.102s1102H}, starquakes could only occur in stars whose initial rotation rate is a substantial fraction of $\Omega_{K}$ (see also \citealt{2018arXiv180404952F}). Assuming full relaxation in each starquake, at most a few such events could occur along the lifetime of the star, much less frequently than observed glitches. Using Eq.~(\ref{recover from glitch 2}), the fractional size of the associated glitches would be
\begin{equation}
\label{von Mises 5}
\frac{\Delta\Omega_{bc}}{\Omega}\approx 2.7 \times 10^{-7}\left(\frac{b}{10^{-5}}\right)\left(\frac{\sigma_c}{0.1}\right)\left(\frac{3}{\tilde\sigma_{M,max}}\right),
\end{equation}
%not much smaller %
within the range of the observed glitches \citep{2000MNRAS.315..534L,2011MNRAS.414.1679E}. Of course, the frequency of the glitches could be increased by assuming either a smaller value of $\sigma_c$ or only partial relaxation in each starquake, but this would make the size of the associated glitches proportionally smaller, so the glitch activity obtained in Eq.~(\ref{glitch activity}) is kept constant.

\

\subsection{Energy released by a starquake}
\label{sec:energy released}

\

\begin{figure}
    \centering
        \resizebox{\hsize}{!}{\includegraphics{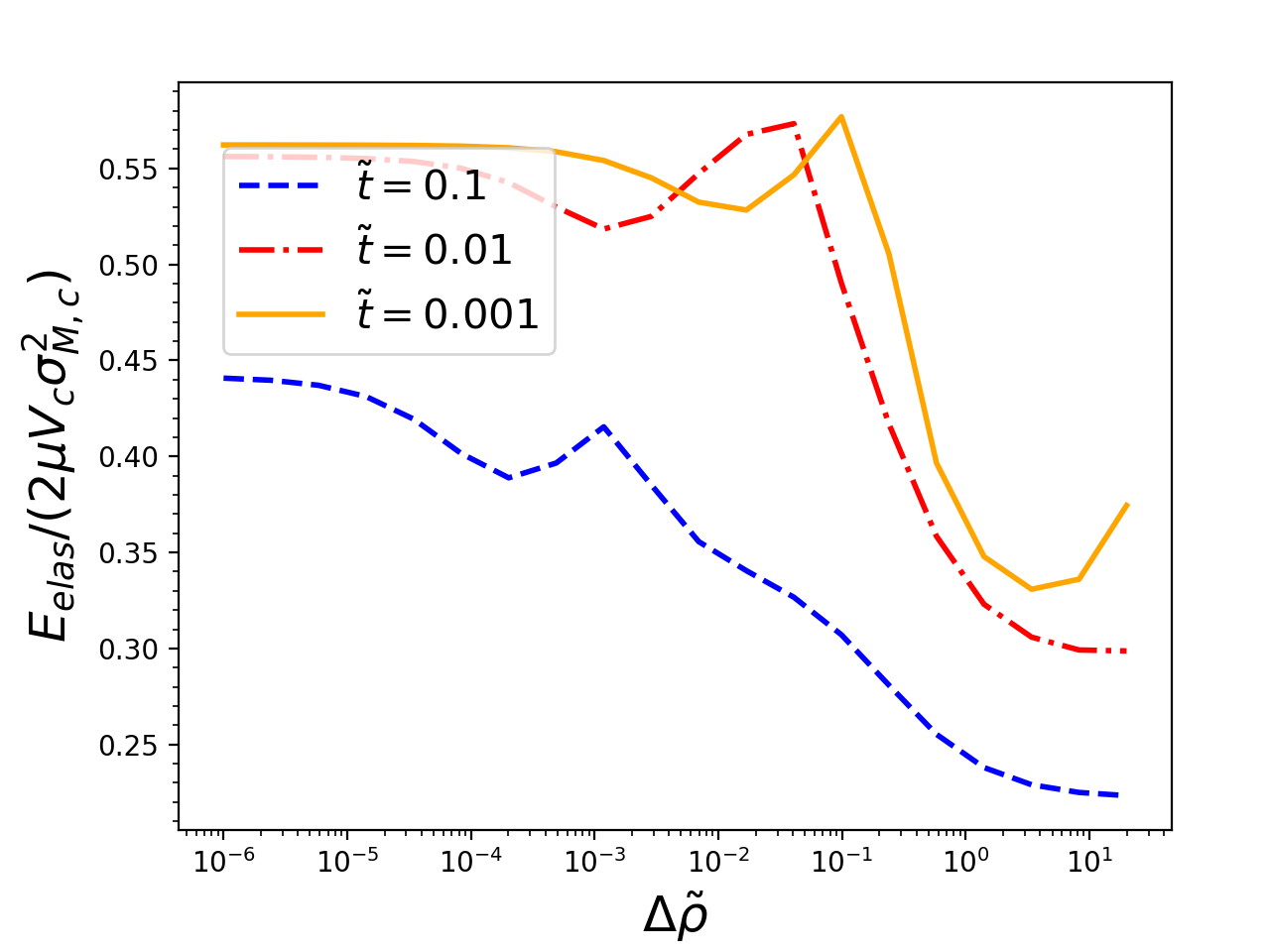}}
    \caption{Elastic energy released when the crust breaks as a function of the density contrast $\Delta\tilde\rho$. It is normalized by the shear modulus $\mu$, crust volume $V_c$, and critical von Mises strain $\sigma_{M,c}$. }
    \label{fig:elastic energy}
\end{figure}

As shown by \citet{1971AnPhy..66..816B}, the changes in the kinetic and gravitational potential energy in a starquake nearly cancel each other, so their sum is much 
smaller than the stored strain energy (roughly by a factor of $b$, as given in Eq.~(\ref{rigidity evaluated})). 
Thus, nearly all of the energy released as the crust breaks comes from the strain energy, 
\begin{equation}
\label{strain energy}
E_{strain}=\int_{crust}\mu \Sigma_{ij}\Sigma^{ij} dV,
\end{equation} 
which can be written as
\begin{eqnarray}\label{strain energy variables}
E_{strain}&=&2\mu V_c\sigma_{M,c}^2\frac{\int_{crust}\tilde\sigma_{M}^2 d (V/V_c)}{ \tilde \sigma^2_{M,max}(\Delta\tilde\rho)} \nonumber \\ 
&\equiv& 2\mu V_c\sigma_{M,c}^2 \tilde E(\Delta\tilde\rho, \tilde r_i),
\end{eqnarray}
with $\tilde E(\Delta\tilde\rho,\tilde r_i)$ shown in Fig. \ref{fig:elastic energy}.
Thus, we can approximate the maximum strain energy released when the crust breaks by rotation as
\begin{equation}
\label{strain_energy_numbers}
E_{strain}\sim 0.6\times 10^{46}\mathrm{erg}\left(\frac{\mu}{10^{30}\mathrm{erg\,cm^{-3}}}\right) 
\left(\frac{\sigma_{M,c}}{0.1}\right)^{2}\left(\frac{V_{crust}}{10^{18}[cm^{3}]}\right) \left(\frac{\tilde E}{0.6}\right). 
\end{equation} 
If all of this energy is deposited as heat into an initially cold star, its final temperature $T_f$ is given by $E_{strain}=\int_0^{T_f}C(T)dT$, where $C(T)$ is the star's heat capacity. Considering only nonsuperfluid neutrons, 
%\begin{equation}
$C(T)%=\frac{\pi^{2}}{2}N k_{B}\frac{T}{T_{F,n}}\sim 2
\approx 1.2\times 10^{39}(T/10^{9}\mathrm{K})\,\mathrm{erg/K}$
(with some dependence on the stellar parameters and the assumed equation of state; \citealt{2017PhRvD..96d3002O}), and the resulting interior temperature is
%. The temperature at which the star is heated after the starquake is
%\begin{equation}
$T_f%=0.7
\approx 1.7\times 10^{8}(E_{strain}/10^{46}\mathrm{erg})^{1/2}\mathrm{K}$.
If, at the opposite extreme, all the neutrons and protons are paired and far below their %pairing, 
respective superfluid or superconducting transition temperatures, the heat capacity is dominated by the electrons. In this case, the heat capacity is reduced by a factor of $\sim 8$, %\citep{2017PhRvD..96d3002O}, %of 10, 
and the final interior temperature is increased to 
%\begin{equation}
$T_f\approx 3\times 10^{8}(E_{strain}/10^{46}\mathrm{erg})^{1/2}\mathrm{K}$. 
These temperatures are in the range where photon luminosity takes over from neutrino luminosity as the main cooling mechanism of neutron stars, which is reached at an age $\sim 10^5\mathrm{yr}$ in standard cooling models (e.g., \citealt{2015SSRv..191..239P}). In younger, hotter neutron stars, the heating due to a starquake is not noticeable, but in older, cooler ones (such as millisecond pulsars) it could have a substantial effect lasting %a time $\sim E_{strain}/L_\gamma
$\sim 10^5\mathrm{yr}$ for each starquake.

It is interesting to compare the energy released in such a starquake with the energy released in the classical model of a glitch due to the sudden coupling transfer of angular momentum from the internal superfluid, of moment of inertia $I_s$ and initial angular velocity $\Omega_s$, to the observable crust and the rest of the star, which rotate together, with moment of inertia $I_c$ and initial angular velocity $\Omega_c<\Omega_s$. 
The total rotational energy of both components before the glitch is $E_{i}=(I_c\Omega_c^2+I_s\Omega_s^2)/2,$ whereas after the glitch both components rotate together, with angular velocity $\Omega_f=(I_c\Omega_c+I_s\Omega_s)/(I_c+I_s)$ and %having a 
rotational energy $E_{f}=(I_c+I_s)\Omega_f^2/2$. Thus, for a glitch of a given observed size $\Delta\Omega_{gl}=\Omega_f-\Omega_c$, the energy released is
\begin{equation}\label{Glitch energy}
E_{gl}\equiv E_i-E_f=\frac{I_c(I_c+I_s)}{2I_s}(\Delta\Omega_{gl})^2, 
\end{equation}
which can be compared to the energy of a starquake of the same size, that is, with $\Delta\Omega_{bc}=\Delta\Omega_{gl}$. For this purpose, we use Eqs.~(\ref{recover from glitch 2}) and (\ref{von Mises 3}) to relate the critical strain and the change in angular velocity in a starquake,
\begin{equation}\label{dOmega starquake}
    \frac{\Delta\Omega_{bc}}{\Omega}=b\Delta\tilde I(\Delta\tilde\rho,\tilde t,0) \frac{\sigma_c}{\tilde\sigma_{max}}.
\end{equation}
With this relation, the ratio of Eqs.~(\ref{Glitch energy}) and (\ref{strain energy variables}) becomes
\begin{eqnarray}
\frac{E_{gl}}{E_{strain}}=\frac{1}{2\tilde E}\left(\frac{\tilde b\Delta\tilde I}{\tilde\sigma_{max}}\right)^2\frac{I_c}{I_s}\frac{\mu V_c E_{rot}}{E_g^2}\sim 10^{-2}\left(\frac{\Omega}{\Omega_K}\right)^2,
%=\frac{\alpha^2\Delta\tilde I^2}{2\tilde E \tilde \sigma_M^2}\left(\frac{I_s}{I_c}\right) \frac{\mu V_c E_{rot}}{E_g^2}
\end{eqnarray}
where we used $\tilde b\equiv b/(\mu V_c/|E_g|)\approx 13$, $I_c/I_s\sim 10^2$, $\mu V_c/|E_g|\sim 10^{-5}$, and $E_{rot}/|E_g|\sim(\Omega/\Omega_K)^2$. We see that for a given change in the rotation rate, the energy released in a starquake (crust yielding) is much larger than in the classical glitch model (superfluid-crust coupling). %So the loss of energy  by transfer of angular momentum by the superfluid component, if the superfluid has $1\%$ of the moment of inertia is $10^{38}[erg]$ for glitches of $\Delta\Omega/\Omega=2.7\times 10^{-6}$, and $10^{32}[ergs]$ for small glitches of $\Delta \Omega/\Omega=10^{-9}$. 

On the other hand, we note that a release of even a very small fraction of the strain energy ($\sim 10^{42}\mathrm{erg}$) could heat the neutron star crust enough to induce a glitch by increasing the mobility of the superfluid neutron vortices \citep{1996ApJ...457..844L,2002MNRAS.333..613L}. Thus, although starquakes by themselves are far from able to explain all the observed glitches (or even the small ones), it might be that starquakes trigger glitches by allowing neutron vortices to move. Since small starquakes releasing only $\sim 10^{-4}$ times the total strain energy are sufficient to cause this effect, they could occur $\sim 10^4$ times more frequently than estimated above, coming close to explaining the total observed glitch activity.

\section{Ellipticity and gravitational wave emission}
\label{sec:ellipticity}

As another application of the calculations presented in the previous sections, we note that rotating neutron stars with a deformation that is not axially symmetric around the axis of rotation emit continuous gravitational waves. Equations (\ref{glitch conservation}) and (\ref{von Mises 5}) yield an explicit expression for $\Delta I_{bc}/I$, a measure of the largest deformation that can be sustained by crustal stresses (and thus released in a starquake) in our model. Of course, this deformation is axially symmetric around the rotation axis ($z$). However, the same deformation could in principle exist for a star that is rotating around an orthogonal axis, say, the $x$-axis. This star would emit gravitational waves of an amplitude proportional to the standard ellipticity,
\begin{equation}\label{ellipticity}
\epsilon\equiv\frac{|I^{yy}-I^{zz}|}{I^{xx}},
\end{equation}
\citep{1998PhRvD..58f3001J}, where $I^{jk}$ are the Cartesian components of the inertia tensor. Defining $\Delta I^{jk}\equiv I^{jk}-I_0$, where $I_0$ is the moment of inertia of the undeformed (spherical) star, it can be shown that $\Delta I^{yy}=-\frac{5}{2}\Delta I^{zz}$, with $\Delta I^{zz}\equiv\Delta I_{bc}$, so
\begin{equation}\label{max ellipticity}
\epsilon_{max}=\frac{7}{2}\frac{|\Delta I_{bc}|}{I}\approx 0.9 \times 10^{-6}\left(\frac{b}{10^{-5}}\right)\left(\frac{\sigma_c}{0.1}\right)\left(\frac{3}{\tilde\sigma_{M,max}}\right).
\end{equation}
In concordance with previous work (e.g., \citealt{2021MNRAS.500.5570G} and references therein), we find this theoretical upper limit to be much higher than observational constraints for millisecond pulsars, $\epsilon\lesssim 10^{-8}$ from the nondetection of gravitational waves \citep{2020ApJ...902L..21A} and in some cases as low as $10^{-9}$ from the small measured slowdown of their rotation (e.g., \citealt{2018ApJ...863L..40W}). Thus, we confirm that the strength of their crust could allow millisecond pulsars to be much less axially symmetric than observed. 
%On the other hand, our result is roughly 40 times larger than the value found for a similar (???) deformation in a different model by \citet{2021MNRAS.500.5570G}. ANALYZE DIFFERENCE BETWEEN THE MODELS!!

\section{Conclusions}
\label{sec:conclusions}

We have studied the possibility of a sudden breaking of the solid crust of a neutron star (``starquake'') as it is stressed by a decreasing rotation rate. In order to do this study in a mostly analytical, parametrized fashion, we considered a Newtonian model in which both the fluid core and the solid crust have uniform densities ($\rho_f$ and $\rho_s$, respectively), whose difference is parametrized by $\Delta\tilde\rho\equiv(\rho_f-\rho_s)/\rho_s$, mimicking the density gradient in a real neutron star. Additional important parameters are $\tilde t\equiv\Delta R/R_\star$ and $\tilde\beta=4\pi\mu R_{\star}^4/(3 G M_{\star}^2)$, where $\Delta R$ is the crust thickness, $R_\star$ is the stellar radius, $M_\star$ is the stellar radius, $G$ is the gravitational constant, and $\mu$ is the crustal shear modulus (also assumed uniform). The conclusions obtained from this model are the following:
\begin{itemize}

    \item The displacement field and shear strain in the crust depend quantitatively and qualitatively on $\Delta\tilde\rho$. If $\Delta\tilde\rho\ll\tilde\beta/\tilde t$ (which includes the case $\Delta\tilde\rho=0$, previously studied by \citealt{2000ApJ...543..987F}), the radial displacement has a maximum at an intermediate radius inside the star, beyond which it decreases toward the surface, and the shape of the crust-core interface does not approach an equipotential in the limit $\mu\to 0$. In the opposite limit, $\Delta\tilde\rho\gg\tilde\beta/\tilde t$, the radial displacement increases monotonically toward the surface, and the core-crust interface does approach an equipotential shape as $\mu\to 0$ (as expected in a fluid star). 
    
    \item Once the shear strain at some point in the crust reaches a critical value $\sigma_c$, the crust breaks, its moment of inertia is suddenly reduced, and an abrupt spin-up occurs, qualitatively similar to the observed pulsar glitches. The size of the glitches caused by this mechanism is proportional to $\sigma_c$, but their frequency is proportional to $1/\sigma_c$. Thus, the resulting glitch activity (sum of glitch sizes over a fixed time interval covering many glitches) is independent of the uncertain value of $\sigma_c$ and turns out to be much smaller than the observed glitch activity in pulsars, even if only the subclass of small glitches is considered. Thus, even this subclass cannot be accounted for by the ``starquake'' model.

    \item For the large values $\sigma_c\sim 0.1$ found in molecular dynamics simulations of the neutron star crust and assuming that each starquake releases most of the strain energy accumulated since the previous event (or since the birth of the star), the associated glitches would be of similar size as their observed counterparts, but they would be extremely rare, happening at most a few times in the life of a neutron star born with a fast initial rotation. Smaller values of $\sigma_c$ or partial relaxation of the crust could account for more frequent but proportionally smaller glitches.
    %Nevertheless, as the change in shape and moment of inertia by spin-down do not explain glitch activity, the neutron star crust can eventually breaks when the critical strain is reached. For a large critical strain of $\sigma_c=0.1$ the change in angular velocity needed to break the crust is really big, so the crust will breaks under spin down a few times in the life of the star.
    
    \item For values of $\Delta\tilde\rho$ in the range $\sim 10-30$ suggested by the average densities of the core and crust in state-of-the-art neutron star models, the ``rigidity parameter'' $b$, namely the ratio between the change of the moment of inertia during a starquake and its change between starquakes, is about an order of magnitude larger than for $\Delta\tilde\rho=0$. %The rigidity parameter we obtain is much larger regarding previous works.  uniform stars all solid, and it is not clear if its depend on the kind of force which is deforming the crust
    
    \item If $\sigma_c\sim 0.1$ and a starquake leads to complete relaxation of the crustal stresses, the energy released can heat an initially cold neutron star up to an internal temperature of a few times $10^8\mathrm{K}$, comparable to the value where thermal photon emission replaces neutrino emission as the main cooling mechanism, at an age $\sim 10^5\mathrm{yr}$. In older neutron stars, such as millisecond pulsars, this would be an appreciable effect.
    
    \item  Even much smaller starquakes could heat the crust enough to cause a large glitch by increasing the mobility of superfluid neutron vortices (e.g., \citealt{1996ApJ...457..844L,2002MNRAS.333..613L}), leaving open the possibility that glitches could be triggered by starquakes. 
    
    \item In our model, the maximum neutron star ellipticity that can be sustained by crustal stresses is $\epsilon_{max}\sim  10^{-6}(\sigma_c/0.1)$, much larger than the constraints $\sim 10^{-8}-10^{-9}$ obtained from timing or gravitational-wave constraints for millisecond pulsars. Thus, we confirm that these stars are much more axially symmetric than required by theory.

\end{itemize}

We stress that our model is quite simplified, assuming that both the core and the crust of the neutron star are uniform, only roughly correcting for gravitational potential perturbations, and not including the effects of general relativity. However, none of these should change our main conclusion, namely that the observed glitches are not starquakes, although even small quakes %the solid crust has far 
could release enough energy to trigger thermally induced glitches. 
%although they could be triggered by the latter. 

\begin{acknowledgements}
We thank Crist\'obal Espinoza, J. Rafael Fuentes, and Luis Rodr{\'\i}guez for providing important data, as well as Enrique Cerda, Curt Cutler, Felipe Espinoza, Elia Giliberti, Charles Horowitz, and Bennett Link for useful discussions and communications. We are also grateful to the anonymous referee for comments that helped improve this paper. This work was funded by FONDECYT Regular Project 1201582. 
\end{acknowledgements}

\bibliographystyle{aasjournal}
\bibliography{1starticle1}

\onecolumn
\begin{appendix}
\section{Coefficients of the radial displacement function}%Functions and parameters of the displacements and shear}

The constants $A_k$ in the radial displacement function (Eq.~(\ref{polynomial})) that satisfy the boundary conditions given in Eqs. (\ref{traction rth interface}), (\ref{boundary condition interface}) 
%at the interface, 
and (\ref{boundary condition surface})
%at the surface, and the vanish of the traction (eq. ) 
are
\begin{eqnarray}
\label{A1}
A_{1}&=& \frac{48}{5\gamma_2}\frac{d_{\Delta1}w_{1}\Delta\tilde{\rho}\tilde{t}+n_{\Delta1}\Delta\tilde{\rho}\beta+n_{1}\beta}{3d_{\Delta1}w_{0}\Delta\tilde{\rho}\tilde{t}+11d_{\Delta2}w_{0} \Delta\tilde{\rho}\beta+11d_{1}\beta+24d_{2}\tilde{t}\beta^{2}},\\
\label{A-2}
A_{-2}&=& \frac{8}{5\gamma_2}\frac{\tilde{r}_{i}^{3}[-3\gamma_2w_{-2}\tilde{t}\Delta\tilde{\rho}+5n_{\Delta2}\Delta\tilde{\rho}\beta+5n_{-2}\beta]}{3d_{\Delta1}w_{0}\Delta\tilde{\rho}\tilde{t}+11d_{\Delta2}w_{0} \Delta\tilde{\rho}\beta+11d_{1}\beta+24d_{2}\tilde{t}\beta^{2}},\\
\label{A3}
A_{3}&=& -\frac{24}{35}\frac{3w_{3}\tilde{t}\Delta\tilde{\rho}+5\tilde{r}_{i}^{5}n_{\Delta3}\Delta\tilde{\rho}\beta+5n_{3}\beta}{3d_{\Delta1}w_{0}\Delta\tilde{\rho}\tilde{t}+11d_{\Delta2}w_{0} \Delta\tilde{\rho}\beta+11d_{1}\beta+24d_{2}\tilde{t}\beta^{2}},\\
\label{A-4}
A_{-4}&=& \frac{3}{35}\frac{\tilde{r}_{i}^{5}[3w_{-4}\tilde{t}\Delta\tilde{\rho}-37\tilde{r}_{i}^{2}n_{\Delta4}\Delta\tilde{\rho}\beta-37n_{-4}\beta]}{3d_{\Delta1}w_{0}\Delta\tilde{\rho}\tilde{t}+11d_{\Delta2}w_{0} \Delta\tilde{\rho}\beta+11d_{1}\beta+24d_{2}\tilde{t}\beta^{2}},
\end{eqnarray}
where 
\begin{eqnarray}
n_{\Delta1}&=& \tilde{r}_{i}^{5}\frac{45+90\tilde{r}_{i}+114\tilde{r}_{i}^{2}+93\tilde{r}_{i}^{3}+72(\tilde{r}_{i}^{4}+\tilde{r}_{i}^{5}+\tilde{r}_{i}^{6})+48\tilde{r}_{i}^{7}+24\tilde{r}_{i}^{8}}{630} \approx 1-\frac{17}{2}\tilde{t},  \\
n_{\Delta2} &=& \tilde{r}_{i}^{2}\frac{19+38\tilde{r}_{i}+57(\tilde{r}_{i}^{2}+\tilde{r}_{i}^{3}+\tilde{r}_{i}^{4})+78\tilde{r}_{i}^{5}+99\tilde{r}_{i}^{6}+80\tilde{r}_{i}^{7}+340\tilde{r}_{i}^{8}}{525} \approx 1-\frac{33}{5}\tilde{t}, \\
n_{\Delta3}&=& %\left(
\frac{8(1+\tilde{r}_{i})+3(\tilde{r}_{i}^{2}+\tilde{r}_{i}^{3}+\tilde{r}_{i}^{4})}%\right)/
{25}  \approx 1-\frac{7}{5}\tilde{t}, \\%  \nonumber , 
n_{\Delta4}&=& %\left(
\frac{19(1+\tilde{r}_{i}+\tilde{r}_{i}^{2})+64(\tilde{r}_{i}^{3}+\tilde{r}_{i}^{4})}
%\right)/
{185}  \approx 1-\frac{101}{37}\tilde{t}, \\
n_0&=&\frac{64(1+\tilde{r}_{i})+43(\tilde{r}_{i}^{2}+\tilde{r}_{i}^{3}+\tilde{r}_{i}^{4})+64(\tilde{r}_{i}^{5}+\tilde{r}_{i}^{6})}{385} \approx 1-3\tilde{t}, \\ 
n_{1} &=& \frac{24+48\tilde{r}_{i}+72(\tilde{r}_{i}^{2}+\tilde{r}_{i}^{3}+\tilde{r}_{i}^{4})+93\tilde{r}_{i}^{5}+114\tilde{r}_{i}^{6}+90\tilde{r}_{i}^{7}+45\tilde{r}_{i}^{8}}{630} \approx 1-\frac{7}{2}\tilde{t}, \\
n_{-2} &=& \frac{40+80\tilde{r}_{i}+99\tilde{r}_{i}^{2}+78\tilde{r}_{i}^{3}+57(\tilde{r}_{i}^{4}+\tilde{r}_{i}^{5}+\tilde{r}_{i}^{6})+38\tilde{r}_{i}^{7}+19\tilde{r}_{i}^{8}}{525} \approx 1-\frac{17}{5}\tilde{t}. \\ 
n_{3}&=& %\left(
\frac{3(1+\tilde{r}_{i}+\tilde{r}_{i}^{2})+8(\tilde{r}_{i}^{3}+\tilde{r}_{i}^{4})}
%\right)/
{25}  \approx 1-\frac{13}{5}\tilde{t}, \\  
n_{-4}&=& %\left(
\frac{64(1+\tilde{r}_{i})+19(\tilde{r}_{i}^{2}+\tilde{r}_{i}^{3}+\tilde{r}_{i}^{4})}
{185} \approx 1-\frac{47}{37}\tilde{t}, \\
d_{\Delta1}&=&\tilde{r}_{i}^{2}\frac{1+2\tilde{r}_{i}+3(\tilde{r}_{i}^{3}+\tilde{r}_{i}^{4}+\tilde{r}_{i}^{5}+\tilde{r}_{i}^{6})+2\tilde{r}_{i}^{7}+\tilde{r}_{i}^{8}}{21} \approx 1-6\tilde{t}, \\
d_{\Delta2}&=& \tilde{r}_{i}^{2}\frac{19(1+\tilde{r}_{i}+\tilde{r}_{i}^{2})+64(\tilde{r}_{i}^{3}+\tilde{r}_{i}^{4}+\tilde{r}_{i}^{5}+\tilde{r}_{i}^{6})+24(\tilde{r}_{i}^{7}+\tilde{r}_{i}^{8}+\tilde{r}_{i}^{9})}{385} \approx 1-\frac{73}{11}\tilde{t},\\ 
d_{1}&=& \frac{24(1+\tilde{r}_{i}+\tilde{r}_{i}^{2})+64(\tilde{r}_{i}^{3}+\tilde{r}_{i}^{4}+\tilde{r}_{i}^{5}+\tilde{r}_{i}^{6})+19(\tilde{r}_{i}^{7}+\tilde{r}_{i}^{8}+\tilde{r}_{i}^{9})}{385} \approx 1-\frac{48}{11}\tilde{t},\\
d_{2}&=& \frac{19+38\tilde{r}_{i}+57\tilde{r}_{i}^{2}+\tilde{r}_{i}^{3}-55\tilde{r}_{i}^{4}+\tilde{r}_{i}^{5}+57\tilde{r}_{i}^{6}+38\tilde{r}_{i}^{7}+19\tilde{r}_{i}^{8}}{175} \approx 1-4\tilde{t},
\end{eqnarray}
and
\begin{equation}
\gamma_{n}=\frac{1+\tilde{r}_{i}+...+\tilde{r}_{i}^{n}}{n+1}  \approx 1-\frac{n}{2}\tilde{t},
\end{equation}
depend only on the dimensionless core radius $\tilde r_i$ (or the complementary crust thickness $\tilde t=1-\tilde r_i$). The coefficients in Eqs.~(A.5) to (A.19) are all defined so that their value is 1 in the limit of a very thin crust ($\tilde t\to 0$), and their expansion up to linear order in $\tilde t$ is given by the approximate expressions at the end of each equation. On the other hand,
\begin{eqnarray}
w_0&=&\frac{g(R_i)}{\tilde{r}_{i}g(R_{\star})}=\frac{\Delta\tilde{\rho}+1}{\tilde{r}_{i}^{3}\Delta\tilde{\rho}+1}, \\
w_1&=&\frac{\gamma_5\Delta\tilde{\rho}+\frac{\gamma_2}{2}}{\tilde{r}_{i}^{3}\Delta\tilde{\rho}+1}, \\
w_{-2}&=& \frac{d_{\Delta1}\Delta\tilde{\rho}}{\tilde{r}_{i}^{3}\Delta\tilde{\rho}+1}, \\
w_{3} &=& \tilde{r}_{i}^{2}\gamma_{2}\frac{\gamma_{7}\Delta\tilde{\rho}+\frac{5}{8}\gamma_{4}}{\tilde{r}_{i}^{3}\Delta\tilde{\rho}+1} \\
w_{-4}&=&\tilde{r}_{i}^{2}\gamma_{2}\frac{\tilde{r}_{i}^{2}\Delta\tilde{\rho}-2\gamma_{1}}{\tilde{r}_{i}^{3}\Delta\tilde{\rho}+1},
\end{eqnarray}
are functions of the thickness of the crust and of the dimensionless density difference $\Delta\tilde\rho\equiv(\rho_f-\rho_s)/\rho_s$.

\section{Explicit expression for the rigidity parameter}

The expression for the rigidity parameter given in Eq.~(\ref{rigidity evaluated}) can be written more explicitly as %, $b \equiv -\Delta I_{bc}/\Delta I_{ac}$, will be given by
\begin{eqnarray}
\label{rigidity evaluated 2}
\lefteqn{
\frac{b}{\mu V_c/|E_g|}%(\Delta\tilde\rho,\tilde t,\tilde\beta)=
% } \\&&\frac{
=[1+\tilde r_i^3\Delta\tilde\rho(\tilde r_i^2+3)/2+\tilde r_i^5\Delta\tilde\rho^2]%} \nonumber \\
%&&\times
\Big\{9\Delta\tilde\rho\tilde t\left(24\eta_2\Delta\tilde\rho^2+16\eta_1\Delta\tilde\rho+3\eta_0\right)+24 d_2 \tilde \beta(1+\tilde r_i^3\Delta\tilde\rho)^2\left[1+(1+\tilde r_i^5)\Delta\tilde\rho+\tilde r_i^8\Delta\tilde\rho^2\right]\Big\}} \nonumber \\
&&\times\Big[5\gamma_2(1+\Delta\tilde\rho \tilde r_i^3)\left(1+(1+\tilde r_i^5)\Delta\tilde\rho+\tilde r_i^8\Delta\tilde\rho^2  \right)  %\nonumber \\
%&&\times
\Big\{\Delta\tilde\rho^3\tilde\beta\tilde r_i^3(11d_{\Delta2}+24\tilde t\tilde r_i^6 d_2 \tilde \beta) %\nonumber \\
%&&
+\Delta\tilde\rho^2[3\tilde t d_{\Delta1}+11(1+\tilde r_i^3)d_{\Delta2}\tilde\beta+11 d_1\tilde r_i^6\tilde\beta+72\tilde r_i^6\tilde t d_2\tilde\beta^2] \nonumber \\
&&+\Delta\tilde\rho(3\tilde t d_{\Delta1}+11d_{\Delta2}\tilde\beta+22\tilde r_i^3d_1\tilde \beta+72\tilde r_i^3\tilde t d_2\tilde \beta^2)+\tilde\beta(11 d_1+24\tilde t d_2\tilde\beta)\Big\} \Big]^{-1}  %\\%^{-1}\nonumber \\
\end{eqnarray}
where $\tilde\beta$ is defined in Eq.~(\ref{beta tilde}), and 
\begin{eqnarray}
\lefteqn{
\eta_0=(19\tilde t^{12} - 285\tilde t^{11} + 1995\tilde t^{10} - 8586\tilde t^9 + 25227\tilde t^8 - 53205\tilde t^7 + 82575\tilde t^6 - 95270\tilde t^5 + 81655\tilde t^4 } \nonumber \\
&& - 51345\tilde t^3 + 22890\tilde t^2 - 6615\tilde t+ 945)/945 \\
\lefteqn{\eta_1= (-19\tilde t^{15} + 342\tilde t^{14} - 2907\tilde t^{13} + 15485\tilde t^{12} - 57855\tilde t^{11} + 160818\tilde t^{10} - 344304\tilde t^9+ 579839\tilde t^8 - 777271\tilde t^7 + 832740\tilde t^6 } \nonumber \\
&&- 710164\tilde t^5 + 475286\tilde t^4 - 242550\tilde t^3 + 89460\tilde t^2 - 21420\tilde t + 2520)/2520
\\
\lefteqn{\eta_2=(19\tilde t^{18} - 399\tilde t^{17} + 3990\tilde t^{16} - 25315\tilde t^{15} + 114525\tilde t^{14} - 393516\tilde t^{13} + 1067608\tilde t^{12} - 2345060\tilde t^{11} + 4237912\tilde t^{10} - 6358740\tilde t^{9} } \nonumber \\
&& + 7947944\tilde t^8 - 8259580\tilde t^7 + 7085460\tilde t^6 - 4952192\tilde t^5 + 2761024\tilde t^4 - 1186920\tilde t^3 + 371280\tilde t^2 - 75600\tilde t + 7560)/7560.
\end{eqnarray}
All these quantities satisfy $\eta_k\to 0$ for $\tilde t\to 1$ (or $\tilde r_i\to 0$), and $\eta_k\to 1$ for $\tilde t\to 0$.

\end{appendix}

\end{document}